\UseRawInputEncoding
\documentclass[final,3p,times,twocolumn]{elsarticle}

\usepackage{amssymb}

\usepackage{amsmath}
\newtheorem{thm}{Theorem}

\newdefinition{rmk}{Remark}
\newproof{pf}{Proof}
\newproof{pot}{Proof of Theorem \ref{thm2}}
\newdefinition{example}{Example}
\newcommand{\tabincell}[2]{\begin{tabular}{@{}#1@{}}#2\end{tabular}}

\usepackage{framed,multirow}
\usepackage{mathtools}
\usepackage{latexsym}
\usepackage{array}
\usepackage{url}
\usepackage{xcolor}
\usepackage{graphicx}
\usepackage{float} 
\usepackage{subfigure} 
\usepackage{booktabs}
\usepackage{stfloats}
\usepackage{appendix}
\usepackage{multirow}
\biboptions{numbers,sort&compress}
\definecolor{newcolor}{rgb}{.8,.349,.1}

\usepackage{lineno}

\begin{document}

\begin{frontmatter}



\title{A new chaotic image encryption algorithm based on transversals \\in a Latin square}

\author[1,2]{Honglian Shen}
\author[1]{Xiuling Shan\corref{cor1}}
\cortext[cor1]{Corresponding author.}
\ead{xiulingshan@sina.com}
\author[1]{Zihong Tian}

\address[1]{School of Mathematical Sciences, Hebei Normal University, Shijiazhuang, 050024, China}
\address[2]{Department of Mathematics and Computer Science, Hengshui College, Hengshui, 053000, China}

\begin{abstract}
There are some good combinatorial structures suitable for image encryption. In this study, a new chaotic image encryption algorithm based on transversals in a Latin square is proposed. By means of an $n$-transversal of a Latin square, we can permutate an image data group by group for the first time, then use two Latin squares for auxiliary diffusion on the basis of a chaotic sequence, finally make use of a pair of orthogonal Latin squares to do the second scrambling. As a whole, the encryption process of ``scrambling-diffusion-scrambling'' is formed. The experimental results indicate that this algorithm achieves a secure and fast encryption effect. The final information entropy is very close to 8,  and the correlation coefficient is approximate to 0. All these tests verify the robustness and practicability of this proposed algorithm.
\end{abstract}



\begin{keyword}
Image encryption \sep Chaotic \sep Latin square \sep Transversals



\end{keyword}

\end{frontmatter}


\section{Introduction}
\label{sec1}

In these years, the network communication develops very rapidly, a large amount of public or private image information is transferred via the public internet. How to transmit a lot of image information safely and efficiently has become an increasingly important issue. Image encryption is the main solution. Digital image encryption is a new branch of computer cryptography, and also a relatively independent branch and a research hot issue in the field of information security. Unlike ordinary text information, digital image has a great deal of data, a strong correlation between pixels and other particularities, which makes DES, IDEA and RSA these traditional methods are not appropriate for them \cite{liao2009}. Therefore, various image encryption algorithms have been put forward in the last few years.

In general, an encryption technology involves two aspects: scrambling and diffusion. Scrambling is mainly used to change the pixels' position, while diffusion is used to change the pixels' value.

Chaos-based encryption algorithms play an important role in existing image encryption algorithms \cite{fridrich1998,wangxy2012novel,belazi2016,xie2017,zhangxiaoqiang2017,wangxy2021novel}. Some qualities  of a chaotic systems like sensitivity to initial values, parameter sensitivity, ergodicity, etc. make it particularly appropriate to do image encryption. Meanwhile chaotic systems also have some disadvantages, such as being defined on a set of real numbers, accompanied by short period phenomena, local linearity and uneven distribution, and requiring discretization when used \cite{lishujun2005}, therefore it's easy to be damaged by chosen plaintext attacks or known plaintext attacks. In addition, many algorithms often use high-dimensional chaotic systems\cite{hua20182d,hua2019cosine,hua2021color,zhangxuncai2021novel,zhoujie2020fast}, increasing the complexity and unpredictability. The higher the dimension of the system,  the more computations are required. Hence, many new different techniques have been used in image encryption algorithms, including one-time keys \cite{liuhj2010color}, bit-level permutation \cite{liuhj2011color,zhou2014,xu2016,xuming2019}, DNA
coding \cite{liuhj2012image,wangxy2015image,chaixiuli2017,dong2021robust}, genetic manipulation \cite{abdullaha2012,zafari2017noise,premkumar2019Gen8,zhangyq2020new}, semi-tensor product theory \cite{wangxy2020for,wangxy2020based}, Fractal sorting matrix \cite{xianyj2021fractal},
compressive sensing \cite{xuqiaoyun2019,khan2021chaos,wangxy2021novel}, and combinatorial designs \cite{wuyue2014compare3,hu2016Design5,xuming2018,xuming2019,zhoujie2020fast,zhangxuncai2021novel,wangxy2021image,hua2021color}.

Recently, many combinatorial design structures have been applied in cryptography, such as Latin square \cite{liguofu2001,wuyue2014compare3,chen2015,hu2016Design5, xuming2018,zhangxuncai2021novel,wangxy2021image}, Latin cube \cite{xuming2019,zhoujie2020fast,hua2021color}, Hadamard matrix \cite{sam2014}, etc. Using simple chaotic system and these uniform combinatorial design structure as auxiliary, better encryption effect can also be achieved. As early as 1949, Shannon pointed out that a perfect password can be expressed by a Latin square in his classic paper \cite{shannon1949}. Kong used Latin squares to build algebraic models for different password systems \cite{colbourn1999}. A Latin square defined on a finite integer set $S$ is a square matrix, and get uniformity for the same number of occurrences of each element in $S$, and the total number of Latin squares is also very large. These characteristics of Latin square are very suitable for image encryption, so some algorithms according to Latin square were put forward. In 2001, Li first proposed a method to do the image scrambling using a pair of orthogonal Latin squares \cite{liguofu2001}, which can directly generate a two-dimensional mapping, instantly increasing the scrambling efficiency. However, he didn't give the algorithm of generating orthogonal Latin squares, nor did he do simulation experiments. In 2014, Wu et al. proposed an image encryption scheme by using Latin square\cite{wuyue2014compare3}. Firstly a Latin square was used to generate a one-dimensional mapping for the scrambling process. Then the Latin square was used to perform bitxor operation on image pixels. However, the scrambling efficiency of this algorithm is low  and it is vulnerable to attack. Other algorithms that use Latin squares have the same problem. In 2018, Xu et al. generated a self-orthogonal Latin square~(SOLS) and proposed a new algorithm for image encryption \cite{xuming2018}. Using a SOLS and its transpose, a pair of orthogonal Latin squares are formed, and the SOLS can provide a pseudo-random sequence for the diffusion process. The experimental results show that this algorithm is safe and highly efficient. The entropy value of encrypted Lena reached 7.997, and the correlation coefficient was small. In 2019, Xu et al. put forward a new image encryption scheme by using 3D bit matrix and orthogonal Latin cubes \cite{xuming2019}. Each original image was decomposed into a three-dimensional bit matrix, and a pair of orthogonal Latin cubes were used not only for confusion but also for diffusion, which proved that the algorithm is highly safe and efficient. In 2020, Zhou et al. combined 3D orthogonal Latin squares (3D-OLSs) with matching matrix for color image encryption \cite{zhoujie2020fast}. The same as algorithm in \cite{xuming2019}, it fully utilized the orthogonality of 3D Latin cube. In 2021, Zhang et al. proposed an encryption algorithm based on Latin square and random shift
\cite{zhangxuncai2021novel}. Wang et al. also put forward an image encryption algorithm based on Latin square array\cite{wangxy2021image}, utilizing the orthogonality and uniformity of Latin square. In 2021, Hua et al. designed a new CIEA using orthogonal Latin squares and 2D-LSM for color image encryption, realized point-to-point permutation and the random distribution of the pixels in the plain image \cite{hua2021color}.

As can be seen from the above discussions, better encryption can also be achieved by using the orthogonality and uniformity of Latin squares, together with other innovations. In addition, there are some other properties of Latin squares that can be utilized, such as transversals. For a Latin square of order $n$, there exists plenty of $n$-transversals. By means of an $n$-transversal, we can divide all $n^2$ positions into $n$ mutually disjoint groups. So in this article we proposed a novel chaos-based image encryption algorithm according to transversals in a Latin square. Fistly we use a $n$-transversal to permutate the image data group by group in the first round of substitution. We can also define two new Latin squares according to the $n$-transversal, which can be used for auxiliary diffusion on the basis of a chaotic sequence and achieve a good diffusion effect. Finally a pair of orthogonal Latin squares are reused for the second scrambling. As a whole, the structure of ``scrambling-diffusion-scrambling'' is formed. Simulation results show that a secure and fast encryption effect is achieved in this algorithm.

In the rest of this article, some primary definitions and conclusions are introduced in Section 2. Section 3 is mainly used to introducing the detailed procedure of encryption and decryption. In Section 4, the experimental results and analysis are demonstrated. In the end, we summarize this article.

\section{Priliminaries}
\label{sec2}

\subsection{Latin squares and transversals}
A Latin square of order $n$ (defined on an $n$-set $S$) is an $n\times n$ array in which each cell contains a single symbol, such that each symbol occurs exactly once in each row and column\cite{xuming2018}. For consistency, we set $S=\{0,1,\ldots,n-1\}$.

Two Latin squares of order $n$ $A=(a_{ij})$ and $B=(b_{ij})$ are orthogonal if every ordered pair $(a_{ij},b_{ij})$ in $S\times S$ occurs exactly once.

Fig.~\ref{1matrix1} lists a pair of orthogonal Latin squares of order 4 $A=(a_{ij})$ and $B=(b_{ij})$. Denote $C=(c_{ij})$ as the juxtaposition array, where $c_{ij}=(a_{ij},b_{ij})$. Each ordered pair in $S\times S$ occurs exactly once.
\begin{figure}[h]
\begin{gather*}
A=
\begin{pmatrix} 0 & 1 &2 & 3\\ 1 & 0 &3 & 2\\2 &3 &0 & 1\\3 & 2 &1 & 0\end{pmatrix}\qquad
B=
\begin{pmatrix} 0 & 1 &2 & 3\\ 2 &3 &0 & 1\\3 & 2 &1 & 0\\1 & 0 &3 & 2 \end{pmatrix}\quad\\
C=
\begin{pmatrix} (0,0) & (1,1) &(2,2) & (3,3)\\ (1,2) & (0,3) &(3,0) & (2,1)\\(2,3) &(3,2) &(0,1) & (1,0)\\(3,1) & (2,0) &(1,3) & (0,2) \end{pmatrix}
\end{gather*}
\caption{Latin squares $A$, $B$ and the juxtaposition array $C$}
\label{1matrix1}
\end{figure}\\
Notation: Using a pair of orthogonal Latin squares $A=(a_{ij})$ and $B=(b_{ij})$ can directly generate a two-dimensional map $\phi:(i,j) \rightarrow  (a_{ij},b_{ij})$,~~$i,j=0,1,...,n-1$. 

Suppose $M$ is a Latin square defined on $S$. A transversal in $M$ is a set of $n$ positions, no two in the same row or column, including each of the $n$ symbols exactly once. Two transversals are disjoint if there are no same positions in them. Any $k$ disjoint transversals are called a $k$-transversal. If $k=n$ there exists an $n$-transversal in $M$.

In Fig.~\ref{1matrix1}, $C$ is the juxtaposition array of $A$ and $B$. Treat each column of $C$ as a position element set of $A$. There are 4 positions in the first column, all row numbers and column numbers are different, and the four elements at the four positions of $A$ are $0,3,1,2$ respectively, so the first column of $C$ is a transversal of $A$. The other columns of $C$ are similar. All the  positions of $A$ are divided into four pairwise disjoint groups, so there is a 4-transversal in $A$.

For an additive group $G$, a bijection $\theta$ of $G$ is called a complete mapping if the mapping $ \sigma:x \xrightarrow\ x+\theta(x)$ is also a bijection of $G$ \cite{anthony2018}.

\begin{thm}\cite{colbourn2007}
\label{th1.2.1}
The Cayley table $M$ of the additive group $G=\{g_{0},g_{1},...,g_{n-1}\}$ is a Latin square with ($i,j$)th entry $g_{i}+g_{j}$. For a bijection $\theta:G \xrightarrow \ G $, $M_{\theta}$ is the Latin square with ($i,j$)th entry $g_{i}+\theta (g_{j})$, and the cells $\{(g_{i}, \theta (g_{i}))~|~i=0,1,...,n-1\}$~~form a transversal of $M$ if and only if $\theta$ is a complete mapping of $G$.
\end{thm}

\begin{thm}
\label{th1.2.2}
 Let $F=\{g_{0},g_{1},...,g_{n-1}\}$ be a finite field with character $p$. $M$ is the Cayley table of $F$. Let $a\in F$, $a\neq 0,1$, and $a\not\equiv -1$ (mod $p$). Define a mapping $\gamma_{j}:x \xrightarrow\ ax+g_{j}$ $(j=0,1,...,n-1)$. Then the following conclusions hold.\\
 1) These $\gamma_{j}$s~($j=0,1,...,n-1$) are $n$ different complete mappings over $F$ under addition;\\
2) Define an $n\times n$ array $M_{\gamma}$ with ($i,j$)th entry $\gamma_{j}~(g_{i})=ag_{i}+g_{j}$. Then $M_{\gamma}$ is a Latin square on $F$;\\
3) Define $D=(d_{ij})$ with $d_{ij}=(g_{i},\gamma_{j}(g_{i}))$. All columns of $D$ form $n$ disjoint transversals of $M$ (named $D$ as the truncated decomposition array). Define the array $M_{1}$ with ($i,j$)th entry $g_{i}+\gamma_{j} (g_{i})$. Then $M,M_{1},M_{\gamma}$ are pairwise orthogonal Latin squares.\
\end{thm}

Appendix A shows the proof of Theorem \ref{th1.2.2}. According to Theorem \ref{th1.2.2}, there are $n$ disjoint transversals in $M$, where the $i$th column index in $j$th transversal is the ($i,j$)th element of $M_{\gamma}$.

\begin{example} \label{1exam1.1}
Let $F$ be a finite field of order 4. Suppose the primitive polynomial is $\omega^{2}+\omega+1$, where $\omega$ is a primitive root of $F$. Let $F=\{g_{0},g_{1},g_{2},g_{3}\}$ with $g_{0}=0$, $g_{1}=1$, $g_{2}=\omega$, $g_{3}=\omega+1$.
\end{example}

Firstly define the Cayley table $M$ of the field $F$ under addition with ($i,j$)th entry $g_{i}+g_{j}$,
\begin{equation*}
M=
\left(
  \begin{array}{cccc}
    0 & 1 & \omega & \omega+1 \\
    1 & 0 & \omega+1 & \omega \\
    \omega &  \omega+1 & 0 & 1 \\
     \omega+1 & \omega & 1 & 0 \\
  \end{array}
\right).
\end{equation*}

Let $a=\omega$. Construct another Latin square $M_{\gamma}$ with ($i,j$)th entry $\gamma _{j}\left ( g_{i} \right )=ag_{i}+g_{j}$,

\begin{equation*}
M_{\gamma}=
\left(
  \begin{array}{cccc}
    0 & 1 & \omega & \omega+1 \\
    \omega &  \omega+1 & 0 & 1 \\
   \omega+1 & \omega & 1 & 0 \\
    1 & 0 & \omega+1 & \omega \\
  \end{array}
\right).
\end{equation*}

Construct the truncated decomposition array $D$ with ($i,j$)th entry $\left ( g_{i},\gamma _{j} \ (g_{i})\right)$,
\begin{equation*}
D=
\left(\setlength{\arraycolsep}{0.48pt}
  \begin{array}{cccc}
    (0,0)\hspace{-0.5cm} & (0,1)\hspace{-0.5cm} & (0,\omega) \hspace{-0.5cm}& (0,\omega+1)\\
     (1,\omega)\hspace{-0.5cm} & (1,\omega+1)\hspace{-0.5cm}& (1,0)\hspace{-0.5cm} & (1,1) \\
    (\omega,\omega+1) \hspace{-0.5cm}& (\omega,\omega)\hspace{-0.5cm} & (\omega,1) \hspace{-0.5cm}& (\omega,0) \\
    (\omega+1,1) \hspace{-0.5cm}& (\omega+1,0) \hspace{-0.5cm}& (\omega+1,\omega+1)\hspace{-0.5cm} & (\omega+1,\omega) \\
  \end{array}
\right).
\end{equation*}
The four positions of each column of $D$ form a transversal of $M$, and the set of all columns is a 4-transversal of $M$.

Finally define the array $M_{1}$ with ($i,j$)th entry $ g_{i}+\gamma _{j} \ (g_{i})=(1+a) g_{i}+g_{j}$,

\begin{equation*}
M_{1}=
\left(
  \begin{array}{cccc}
    0 & 1 & \omega & \omega+1 \\
   \omega+1 & \omega & 1 & 0  \\
    1 & 0 & \omega+1 & \omega\\
    \omega &  \omega+1 & 0 & 1  \\
  \end{array}
\right).
\end{equation*}

According to Theorem \ref{th1.2.2}, $M,M_{1},M_{\gamma}$ are pairwise orthogonal Latin squares.

\subsection{Logistic map}
In this article, we adopt the classical Logistic map to genrate two new sequences. One of them is used to generate a finite field, and the other is used to do diffusion. We describe Logistic map as follows.
\begin{equation}
\label{logisticmap}
x_{i+1}=\lambda x_{i}(1-x_{i}),  ~~~~i=0,1,2,...
\end{equation}
where $\lambda$ is a system parameter, $0 < \lambda \leqslant 4$, and $x_{i} \in \left ( 0,1 \right )$. When $\lambda >3.573815$, the sequence shows chaos.

\section{The proposed image encryption algorithm}
\label{sec3}
\begin{figure}[!htb]
\centering
\includegraphics[width=3in]{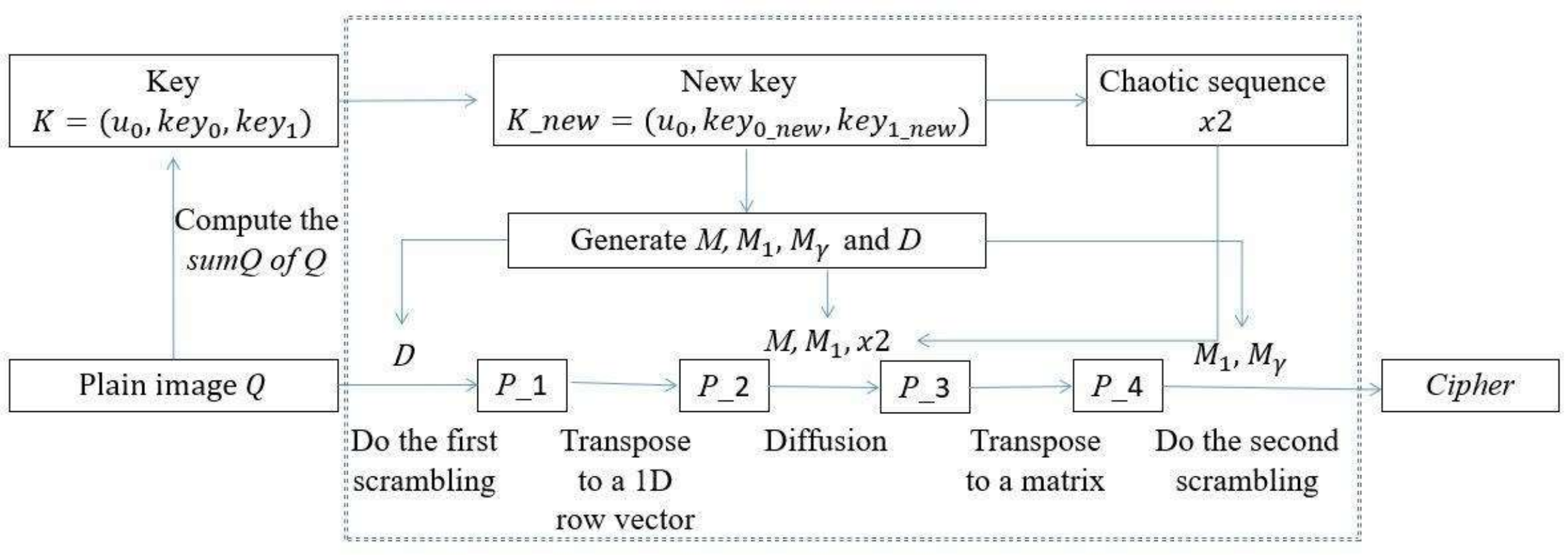}
\caption{The encryption process}
\label{1encryprocess}
\end{figure}

For simplicity, some of the symbols are described as follows. $n$ stands for a prime power. $Q$ is used to represent an $n \times n$ original plaintext image, $K$ is the encryption key and $Cipher$ denotes the corresponding ciphertext. This algorithm is divided into two parts: Algorithm 1 generates three Latin squares and an $n$-transversal by use of $K$ and the features of $Q$; Algorithm 2 is mainly used for encryption, including three layers: scrambling, diffusion and scrambling, then the encrypted image $Cipher$ is formed. The encryption diagram is listed in Fig.~\ref{1encryprocess}.
\subsection{The generation of Latin squares $M,M_{1},M_{\gamma}$ and an $n$-transversal}
We use Algorithm 1 to construct three Latin squares and an $n$-transversal, all of which are directly generated on a finite field, using addition and multiplication in the finite field.\\
\\
Algorithm 1:~The generation of $M,M_{1},M_{\gamma}$ and an $n$-transversal

Input: An $n \times n$ plain image $Q$, encryption key $K=(\mu_{0},key_{0},key_{1})$, public parameter $a$.

Output: Latin squares $M,M_{1},M_{\gamma}$ and the truncated decomposition array $D$. \\

Step~1:~Compute the sum of all pixels in $Q$, denoted as $sumQ$. Let
\begin{equation}
\label{1formula_s}
s=floor(sumQ/255 \times 10^{15})/10^{15},
\end{equation}
where floor is the downward integer function. Compute $key_{0\_new}=(key_{0}+s)/2$, $key_{1\_new}=(key_{1}+s)/2$. It's very essential because $sumQ$ reflects the characteristics of the plaintext image. When the plaintext image changes a little, the chaotic sequence will change greatly because of the changed key. In other words, only one round of encryption is needed to achieve a high sensitivity to the plaintext image.

Step~2:~Generate a logistic sequence of length $n$ $x1=\{x_{i}~|~i=0,1,2,...,n-1\}$ with system parameter $\mu _{0}$ and initial value $x_{0}=key_{0\_new}$. Sort $x1$ as follows:\
\begin{equation}
[fx,lx]=sort(x1)\\
\end{equation}
where $sort$ is the function that sorts a sequence in ascending order. $fx$ is the new sequence reordered by $x1$, $lx$ is the index position.

Step~3:~Redefine operations of addition and multiplication on $lx$, then generate a finite field $F_{n}$ with character $p$. Denote $F_{n}=\{g_{0},g_{1},...,g_{n-1}\}$. Select $a\in F_{n}$, $a\neq 0,1$, and $a\not\equiv -1$(mod~$p$), generate three Latin squares $M,M_{1},M_{\gamma}$ with ($i,j$)th entry $g_{i}+g_{j}$, $(1+a)g_{i}+g_{j}$ and $ag_{i}+g_{j}$ respectively. According to Theorem \ref{th1.2.2}, $M,M_{1},M_{\gamma}$ are pairwise orthogonal.

Step~4:~Generate the truncated decomposition array $D$ with ($i,j$)th entry ($g_{i},ag_{i}+g_{j}$). Then the column set of $D$ is an $n$-transversal of $M$.

\subsection{Image encryption}

We use Algorithm 2 to complete the rest of the encryption process. First of all, with the help of the truncated decomposition array $D$, we can permutate the image data $Q$ group by group. Secondly, we use two Latin squares $M$ and $M_{1}$ for auxiliary diffusion based on another chaotic sequence $x2$. Finally, a pair of orthogonal Latin squares  $M_{1}$ and $M_{\gamma}$ are used for the second scrambling. The following is the detailed procedure of Algorithm 2.\\
\\
Algorithm 2: The proposed encryption algorithm

Input: An $n \times n$ plain image $Q$, encryption key $K=(\mu_{0},key_{0},key_{1})$, public parameters $a$, $c_{1}$ and $c_{2}$.

Output: Ciphertext image $Cipher$. \\

Step~1:~Make use of Algorithm 1, $Q,K$ and $a$ to generate $M,M_{1},M_{\gamma}$ and $D$.

Step~2:~Scramble $Q$ for the first time. At first convert $D$ to a natural column index array $D_{\theta}$ by bijection $\theta: g_{i}\rightarrow i$. Starting from the first transversal, the first pixel of $Q$ at $D_{\theta}(0,0)$ is placed at the position $D_{\theta}(1,0)$, the second pixel at the position $D_{\theta}(2,0)$ is placed at the position $D_{\theta}(3,0)$, and so on..., until the last pixel at the position $D_{\theta}(n-1,0)$ is placed at the position $D_{\theta}(0,0)$. After scrambling $n$ times based on $n$ transversals, we can get a temporary image $P\_1$. The specific process is shown below.
\begin{eqnarray}
\left\{
\begin{aligned}
&P\_1(D_{\theta}(i+1,j))=Q(D_{\theta}(i,j)),\\
&P\_1(D_{\theta}(0,j))=Q(D_{\theta}(n-1,j)),\\
&0\leq i \leq n-2,~~0\leq j \leq n-1.\\
\end{aligned}
\right.
\end{eqnarray}
%
%
%
%
%

\begin{figure*}[!htb]
\centering
\subfigure[]{
\label{1M_jietai}
\includegraphics[width=0.53\textwidth]{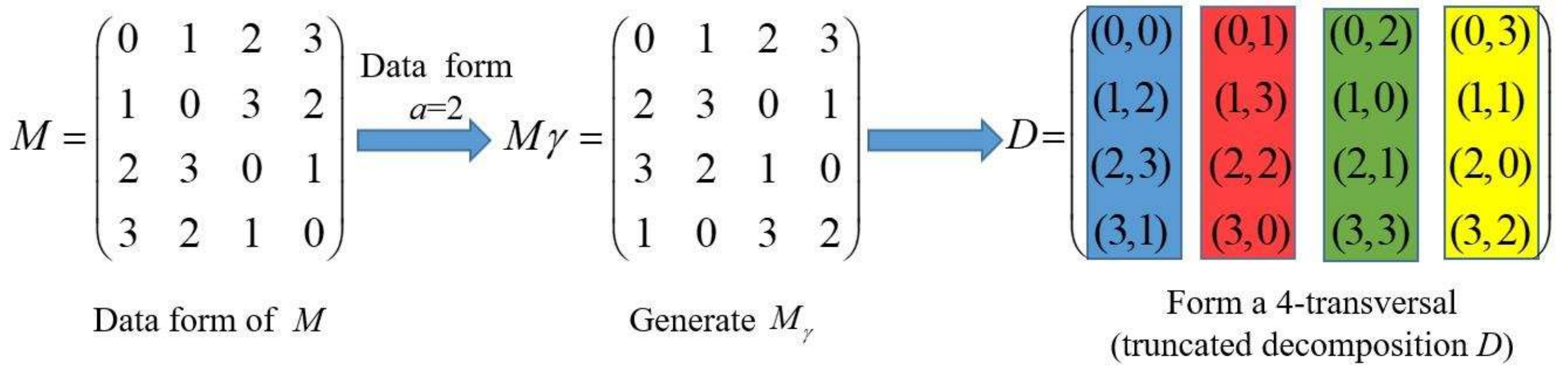}}
\hspace{0.4cm}
\subfigure[]{
\label{1Q_jietai}
\includegraphics[width=0.4\textwidth]{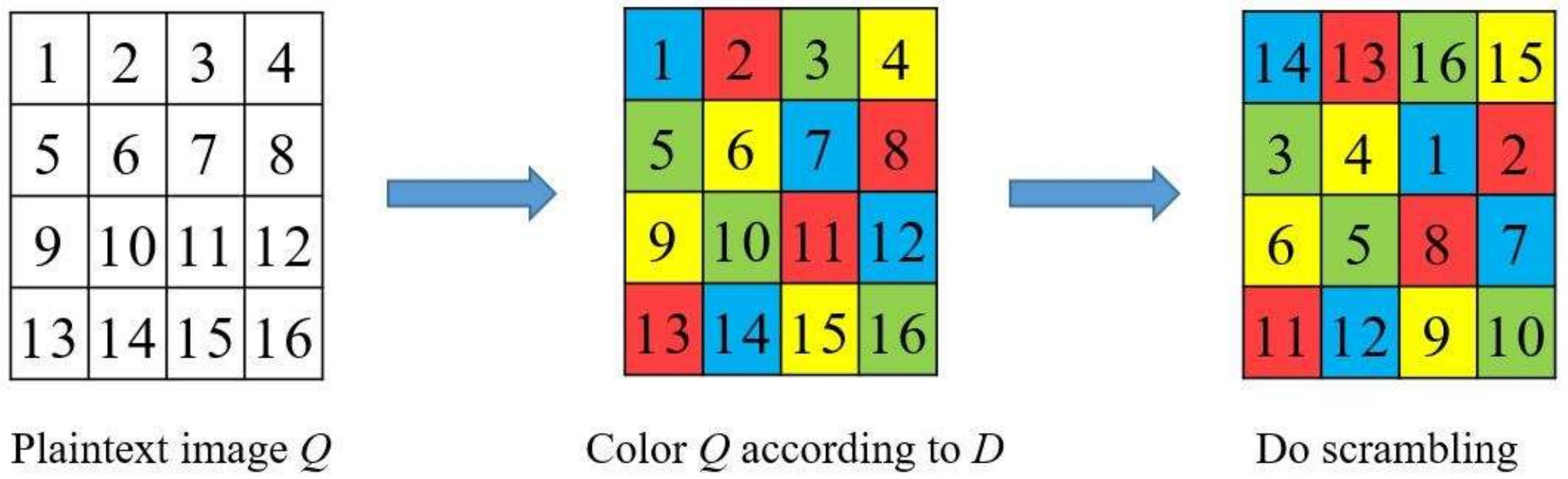}}
\caption{An example of doing scrambling according to a 4-transversal}
\label{1jietaizhiluan}
\end{figure*}

Fig.~\ref{1jietaizhiluan} shows a 4-order example to illustrate the scrambling process in this step. In Fig.~\ref{1jietaizhiluan}(a), A Latin square $M$(Generated on the field of Example \ref{1exam1.1}) is converted to digital form. Select a number $a=2$, then generate $M_{\gamma}$ with the $(i,j)$th entry $(1+a)g_{i}+g_{j}$, resulting in a 4-transversal $D$, all the 16 positions of a 4-order matrix are divided into four pairwise disjoint groups. Because $g_{i}=i$, $D_{\theta}=D$, we can scramble $Q$ according to $D$. In Fig.~\ref{1jietaizhiluan}(b), starting from the first transversal, the first pixel ‘1’ at (0,0) is placed at (1,2), the second pixel ‘7’ at (1,2) is placed at (2,3), the third pixel ‘12’ at (2,3) is placed at (3,1), finally, the fourth pixel ‘14’ at (3,1) is placed at (0,0). So are the scramblings of the other transversals.
Because $D$ is a 4-transversal, the first scrambling can be completed after 4 times .

Step~3:~Firstly convert $P\_1$ into a row vector $P\_2$, then generate another new chaotic sequence of length $n^2+100$ with system parameter $\mu_{0}$ and initial value $key_{1\_new}$. To eliminate the effect of the initial value, especially delete the first 100 digits, the rest form a new chaotic sequence $x2$. $M$ and $M_{1}$ are transposed into row vectors $L_{M}$ and $L_{M1}$, which are used as two pseudo-random sequences for auxiliary diffusion to form a new row vector $\{P\_3(i)\}_{i=0}^{n^{2}-1}$. The detailed diffusion formula is as follows.
\begin{eqnarray}
\left\{
\begin{aligned}
&b=mod(floor(x2(i)*(10^3+c_{1}*L_{M}(i)\\
&~~~~+c_{2}*L_{M1}(i))),256),\\
&P\_3(i)=P\_2(i)\oplus b\oplus P\_3(i-1).\\
\end{aligned}
\right.
\end{eqnarray}
where the initial value $P\_3(-1)=0$, $b$ is a temporary variable, and mod is the module integer function.

Step~4:~Transpose $P\_3$ to an array $P\_4$. By using the orthogonality of $M_{1}$ and $M_{\gamma}$ we conduct the second scrambling by formula (\ref{1formula_scram2}), the final ciphertext image $Cipher$ is obtained.
\begin{eqnarray}
\left\{
\begin{aligned}
  \label{1formula_scram2}
&P\_4(M_{1}(i,j), M_{\gamma}(i,j))\rightarrow Cipher(i,j),\\
& 0\leq i,j \leq n-1.\\
\end{aligned}
\right.
\end{eqnarray}

\subsection{Image decryption}
When we do image decryption, follow the reverse procedure, and we need to know the value $sumQ$ in advance. The following figure is the decryption diagram.
\begin{figure}[!htb]
\centering
\includegraphics[width=3in]{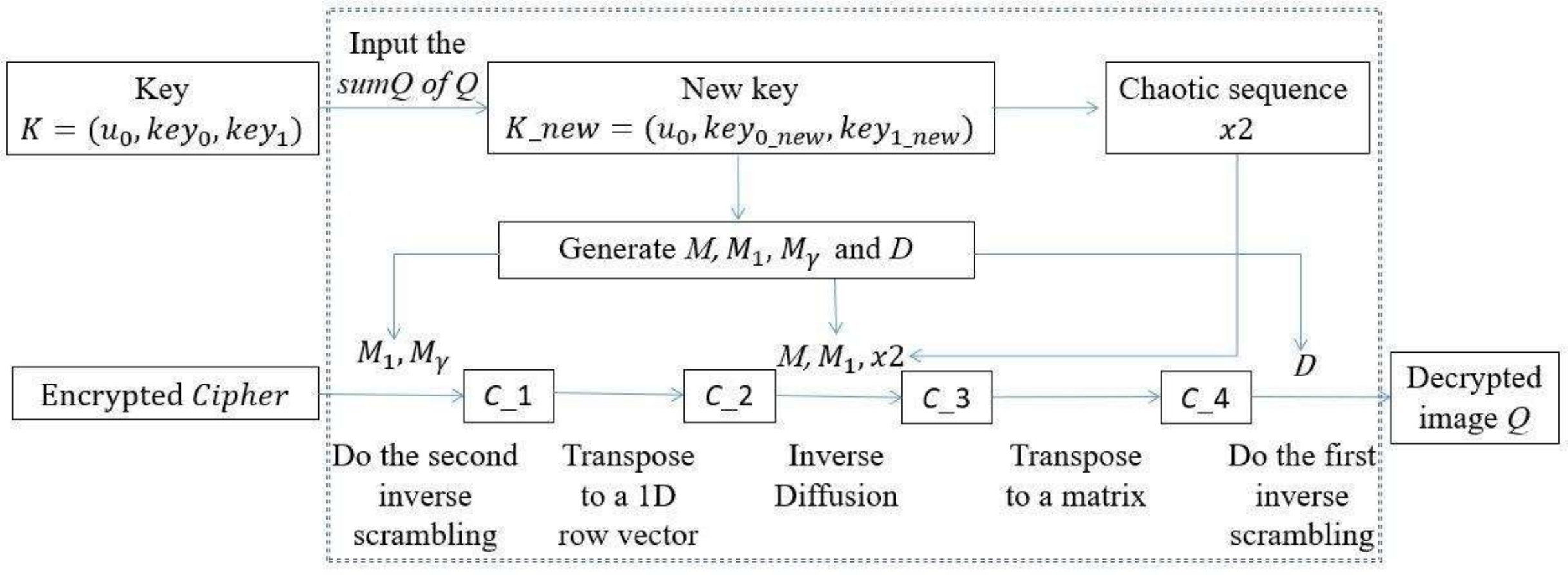}
\caption{The decryption process}
\label{1decryprocess}
\end{figure}

\section{Simulation results and security analysis}
We conducted simulation experiments and listed all the results in this section. In order to reflect the advantages of this algorithm, we compared this algorithm with some representative algorithms \cite{wangxy2021image,liuhui2019Qua4,xuming2018,xu2017,belazi2016,zhangxuncai2021novel,cao2018novel}.

In our experiments, a total of six different $256 \times 256$ images are selected for testing, which are chosen from the USC-SIPI2 and CVG-UGR3 image sets. Every experiment requires only one round of encryption, the secret key $K$ is: $\mu_{0}$= 3.99999, $key_{0}$=0.123456, $key_{1}$=0.234567. There are three public parameters $a=\omega$~($\omega$ is a primitive root of $F_{n}$), $c_{1}=1.3, c_{2}=1.5$.

The algorithm was tested from the following aspects: key space and sensitivity analysis, histogram test, correlation test, information entropy test, differential attack resistance test, robustness test, computational complexity and time efficiency analysis, and so on.
\subsection{Key space and sensitivity analysis }
\subsubsection{Key space analysis}
There are three real numbers in $K=(\mu_{0},key_{0},\\key_{1})$, and the computational accuracy of each value is $10^{-15}$,  so this algorithm can achieve a key space of $10^{45}\approx 2^{149}$, greater than $2^{128}$ \cite{jincong2017key1,kulsoom2016key2,patro2018key3}. There are also three public parameters to select, so the algorithm has a large enough key space and occupies a relatively small memory space. In summary, it can resist brute-force attacks.
\subsubsection{Key sensitivity analysis}
An excellent image encryption algorithm desires strong sensitivity to the key, so sensitivity analysis is often considered as a crucial indicator of resistance to brute-force attacks. It's usually evaluated from two aspects: sensitivity during encryption and sensitivity during decryption.

(1) Key sensitivity analysis during encryption

\begin{figure}[!htb]
\centering  
\subfigure[Two ciphers with $\mu_{0}$ changed $10^{-15}$]{
\label{1encry_u0}
\includegraphics[width=0.45\textwidth]{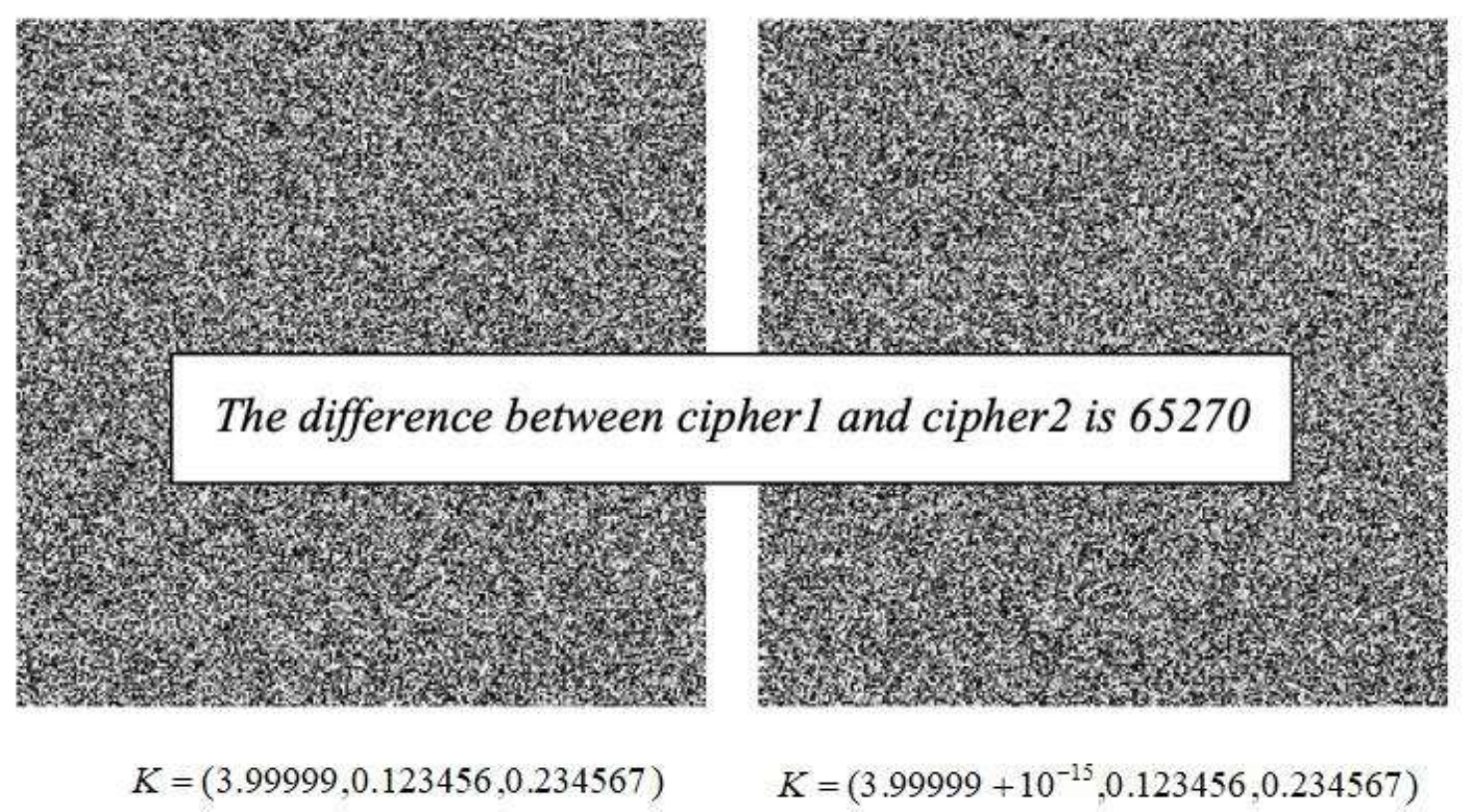}}
\subfigure[Two ciphers with $key_{0}$ changed $10^{-15}$]{
\label{1encry_key0}
\includegraphics[width=0.45\textwidth]{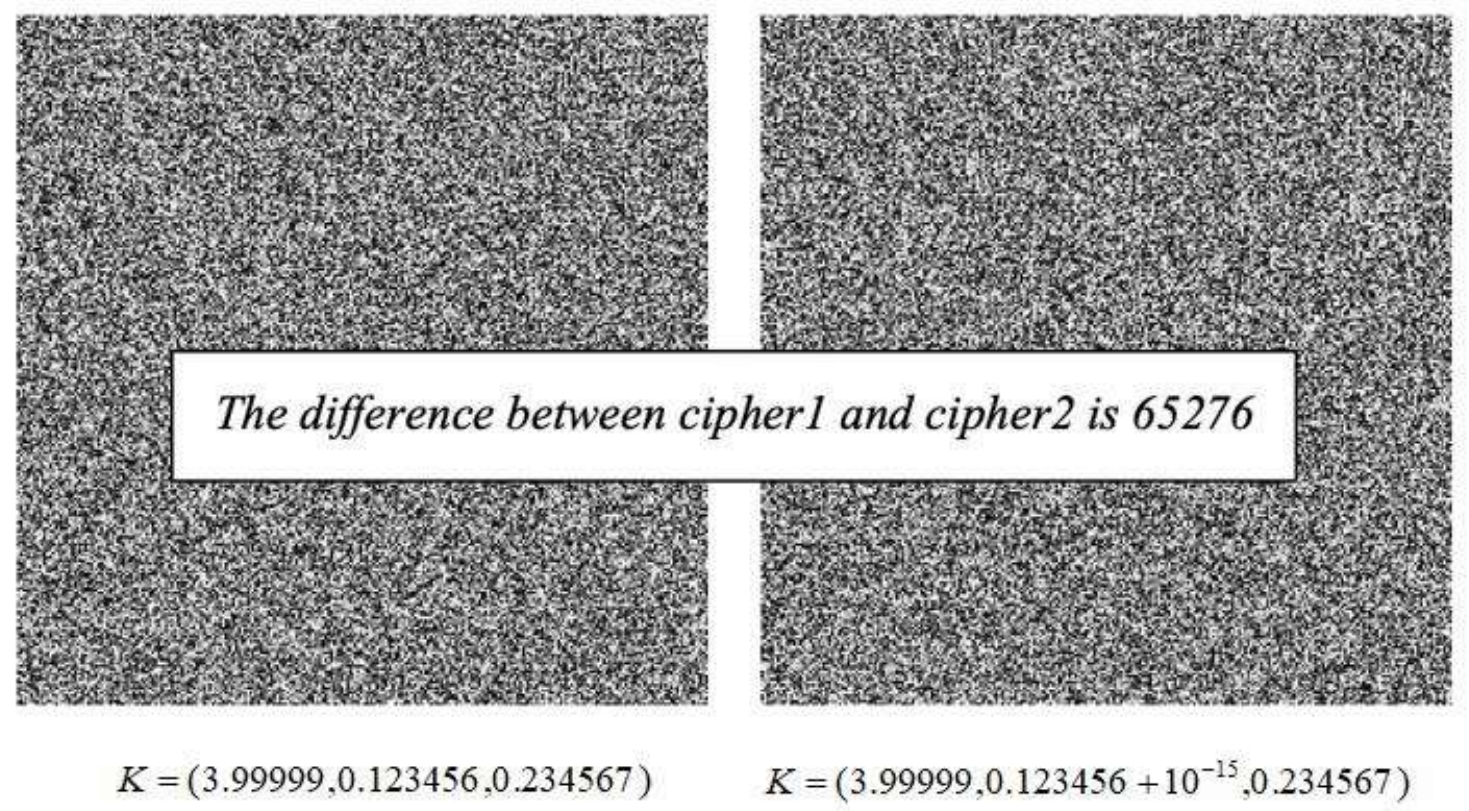}}
\subfigure[Two ciphers with $key_{1}$ changed $10^{-15}$]{
\label{1encry_key1}
\includegraphics[width=0.45\textwidth]{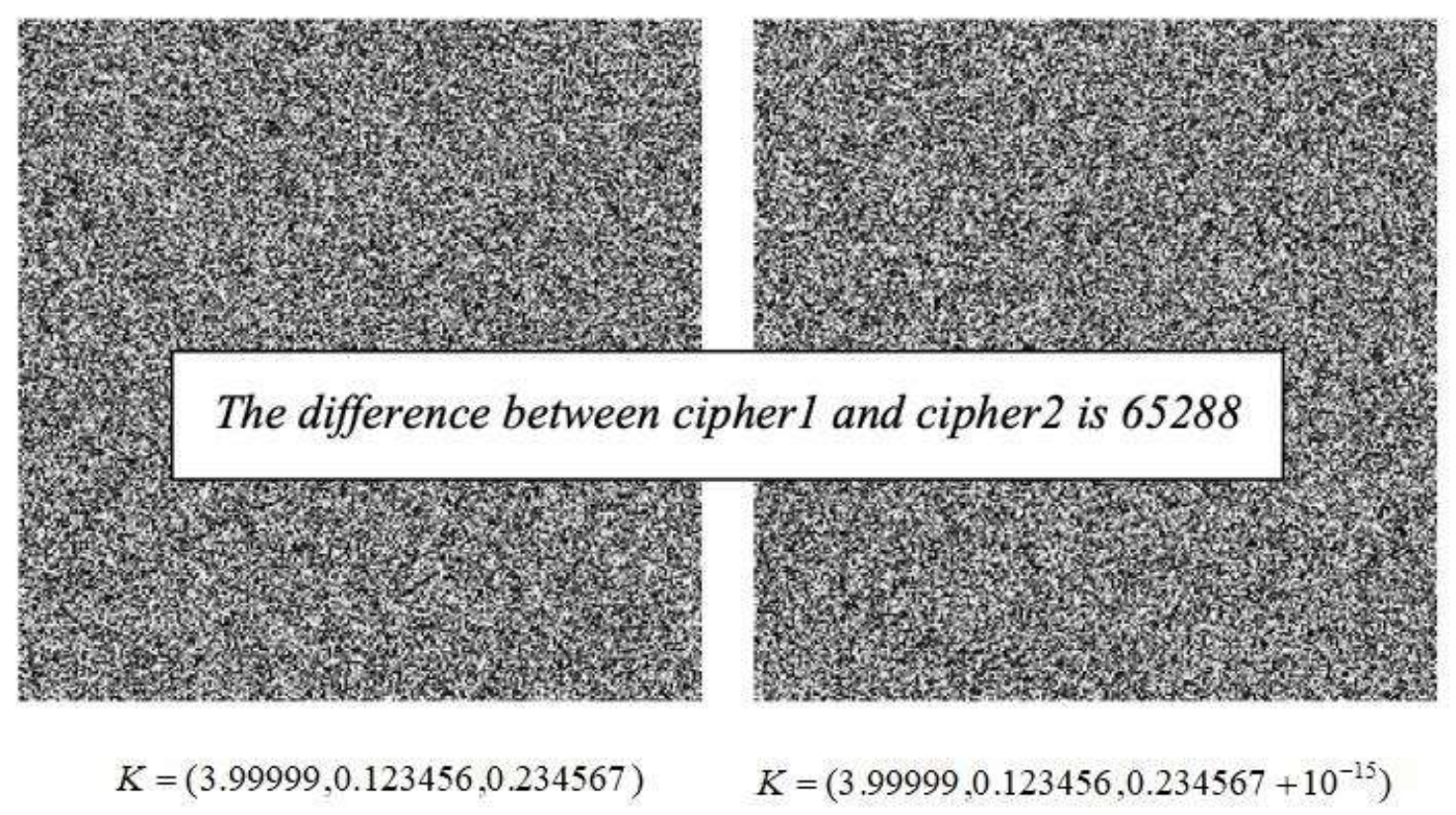}}
\caption{Comparisons of encryption results with key changed}
\label{1encry_sensitive}
\end{figure}

Take Lena for example. Firstly, set $K=(3.99999,0.123456,0.234567)$, then modify each value slightly by adding  $10^{-15}$ after the decimal point. We use two sets of secret keys to encrypt Lena, $Cipher1$ being the image encrypted with $K$ and $Cipher2$ being the image encrypted with the modified $K$.
Fig.~\ref{1encry_sensitive} shows the comparison of the results of the two ciphertext images. Percentages of different pixels are computed as shown in Table \ref{1table1_encry}, which are all greater than 99.59\%, fully indicating that the algorithm is extremely sensitive to the key during encryption.
\begin{table}[!htb]
\setlength{\abovecaptionskip}{0pt}
\caption{Key sensitivity test results during encryption}
\label{1table1_encry}
\centering
\scriptsize
\scalebox{0.9}{
\begin{tabular}{llll}
\toprule
The comparison ciphers & Fig.~\ref{1encry_u0} & Fig.~\ref{1encry_key0} & Fig~.\ref{1encry_key1}\\
\midrule
Number of different pixels & 65270 & 65276 & 65288 \\
percentage & 99.5941\% & 99.6033\% & 99.6216\%  \\
\bottomrule
\end{tabular}}
\end{table}

(2) Key sensitivity analysis during decryption

\begin{figure}[!htb]
\centering  
\subfigure[Two deciphers with $\mu_{0}$ changed $10^{-15}$]{
\label{1decry_u0}
\includegraphics[width=0.45\textwidth]{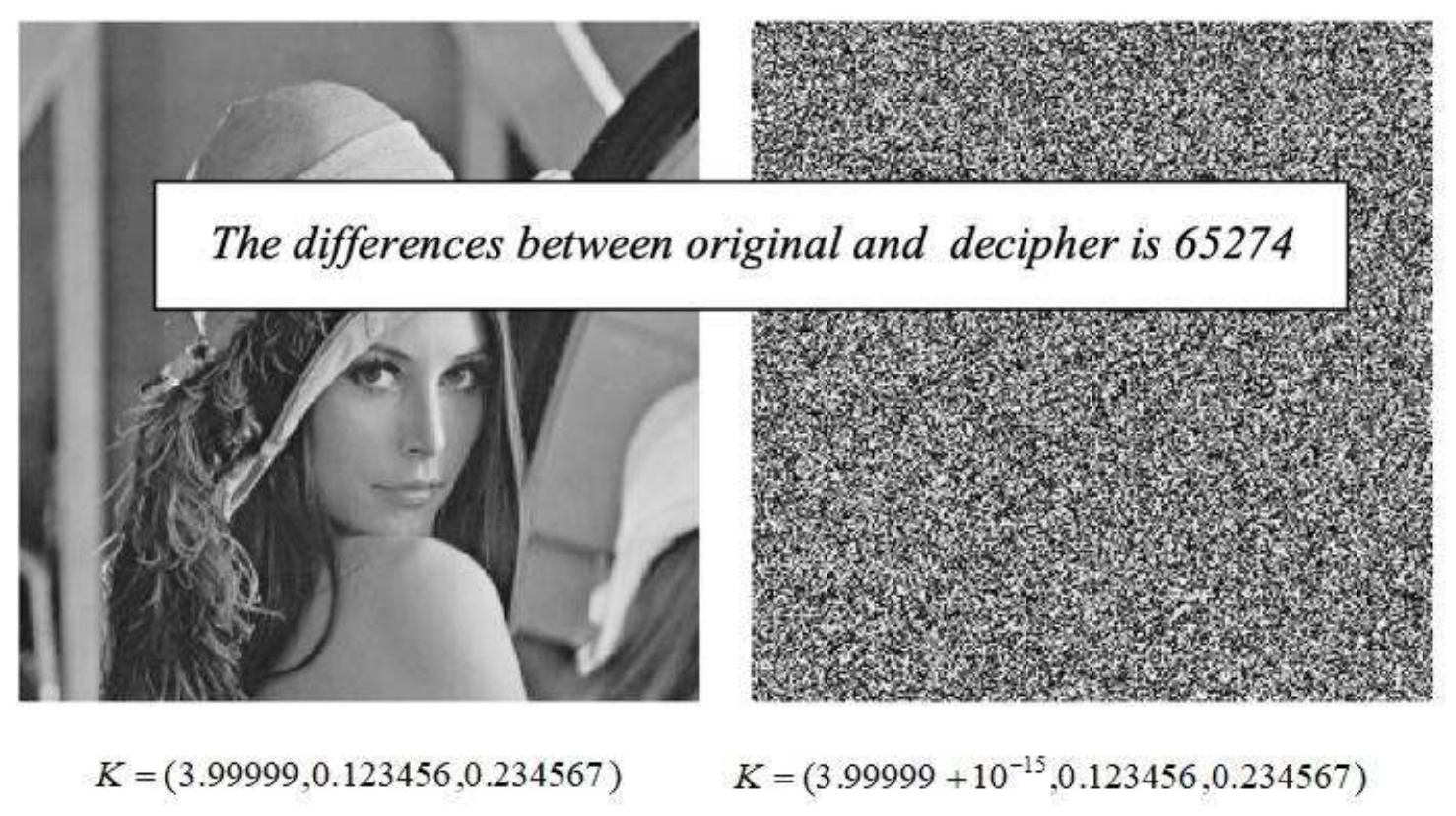}}
\subfigure[Two deciphers with $key_{0}$ changed $10^{-15}$]{
\label{1decry_key0}
\includegraphics[width=0.45\textwidth]{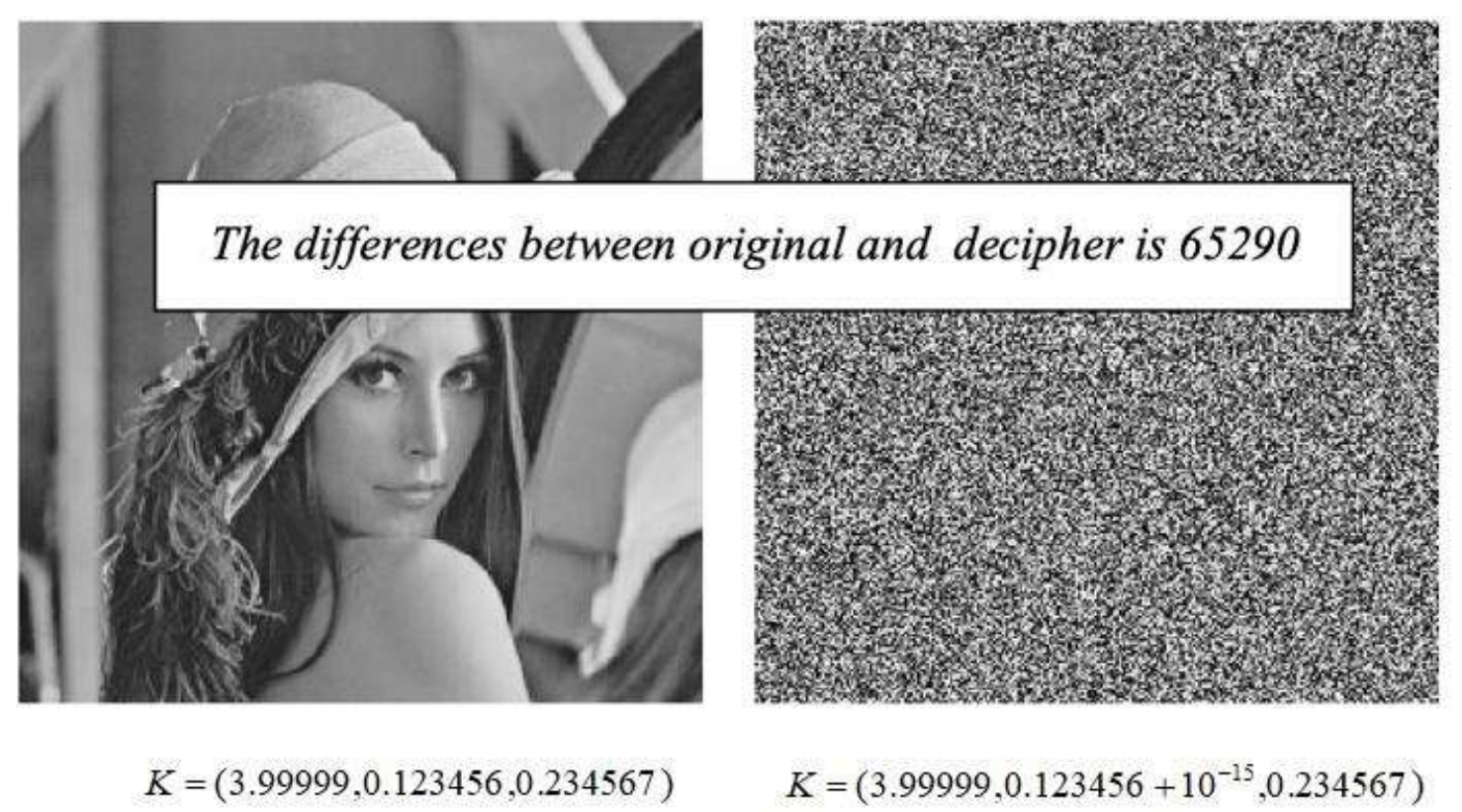}}
\subfigure[Two deciphers with $key_{1}$ changed $10^{-15}$]{
\label{1decry_key1}
\includegraphics[width=0.45\textwidth]{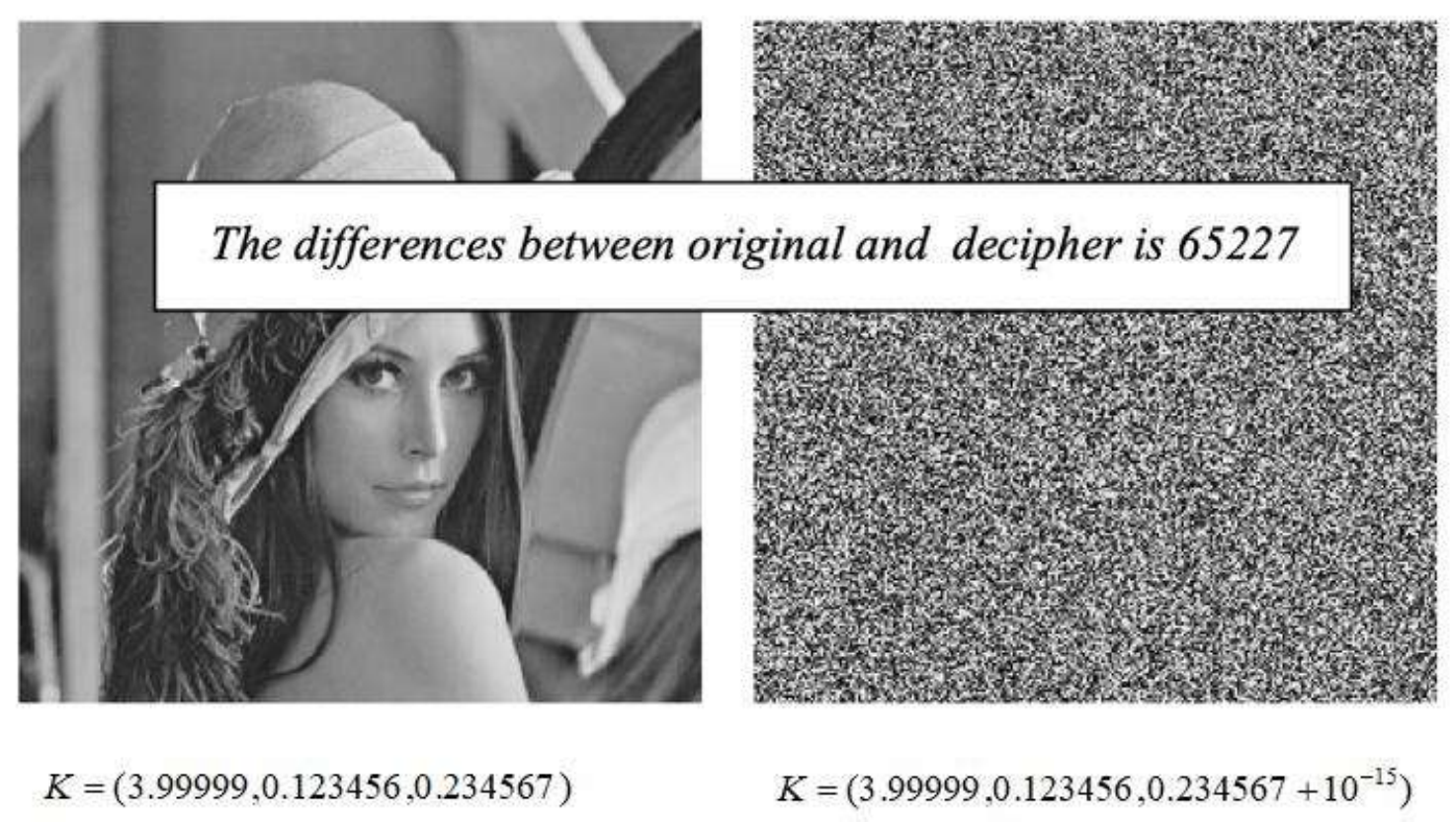}}
\caption{Comparisons of decryption results with key changed}
\label{1decry_sensitive}
\end{figure}

Similarly, Lena is also used to do key sensitivity analysis during decryption. In view of the encrypted image $Cipher$, make a tiny change value $10^{-15}$ in each value of $K=(3.99999,0.123456,0.234567)$, then use the two sets of secret keys to decrypt $Cipher$. Fig.~\ref{1decry_sensitive} presents when we use the right key, we can obtain the original image, while using the modified key, we can't decrypt correctly. In addition, Table \ref{1table2_decry} records the percentages of different pixels of two deciphers, all greater than 99.5\%. From these results, we can discover that even though the key changes a little, we will fail to obtain the original image. So this algorithm is key-sensitive during decryption.
\begin{table}[!htb]
\setlength{\abovecaptionskip}{0pt}
\caption{Key sensitivity test results during decryption}
\label{1table2_decry}
\centering
\scriptsize
\scalebox{0.9}{
\begin{tabular}{llll}
\toprule
Original and decrypted image & Fig.~\ref{1decry_u0} & Fig.~\ref{1decry_key0} & Fig.~\ref{1decry_key1}\\
\midrule
Number of different pixels & 65274 & 65290 & 65227 \\
percentage & 99.6002\% & 99.6246\% & 99.5285\%  \\
\bottomrule
\end{tabular}}
\end{table}

\subsection{Statistical Analysis}
A good algorithm for image encryption should be capable of resisting any statistical attacks. The main statistical indicators include histogram analysis, correlation coefficients of adjacent pixels (usually considering three directions) and information entropy analysis.
\subsubsection{Histogram analysis}
In an image, histogram is a representation of the frequency of each gray level pixel. A well-encrypted image has a histogram distribution that is as uniform as possible. In general, it can be measured by variance $S$, and the formula is as follows:
\begin{equation}
S=\frac{1}{256}\sum_{i=0}^{255}(hist_{i}-aver)^{2}.
\end{equation}
where $hist_{i}$ denotes the frequency of the $i$th gray level pixel, $aver=\frac{1}{256}\sum_{i=0}^{255}hist_{i}$. $S$ represents the variance of histogram. The smaller the value of $S$, the better.

\begin{figure*}[!htb]
\centering  
\includegraphics[width=0.9\textwidth]{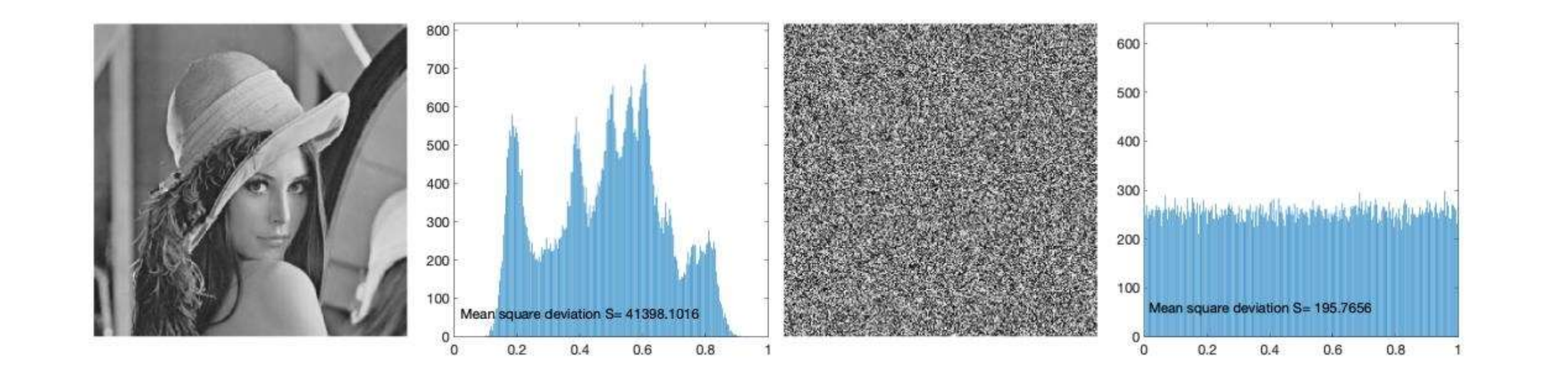}
\includegraphics[width=0.9\textwidth]{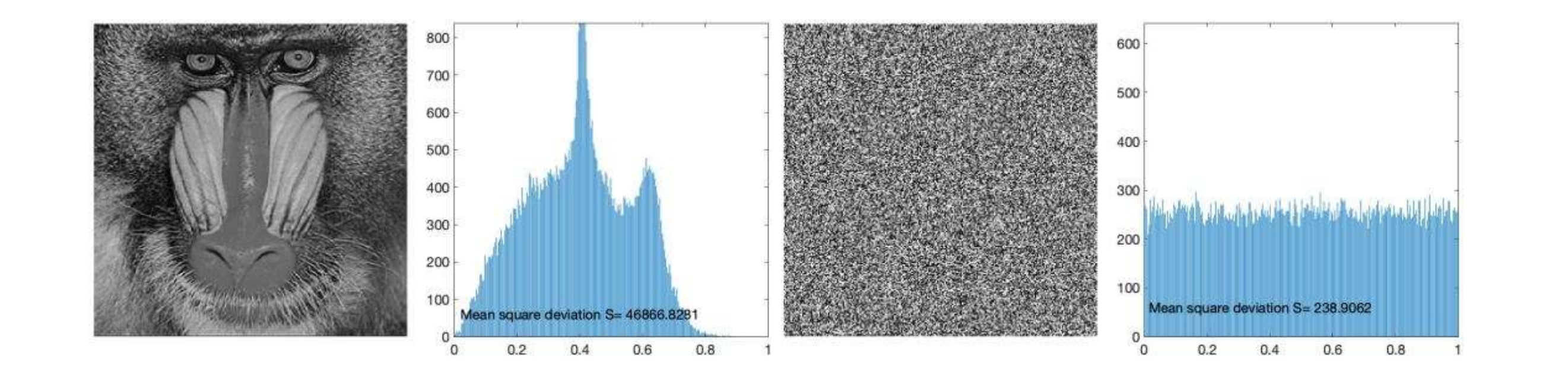}
\includegraphics[width=0.9\textwidth]{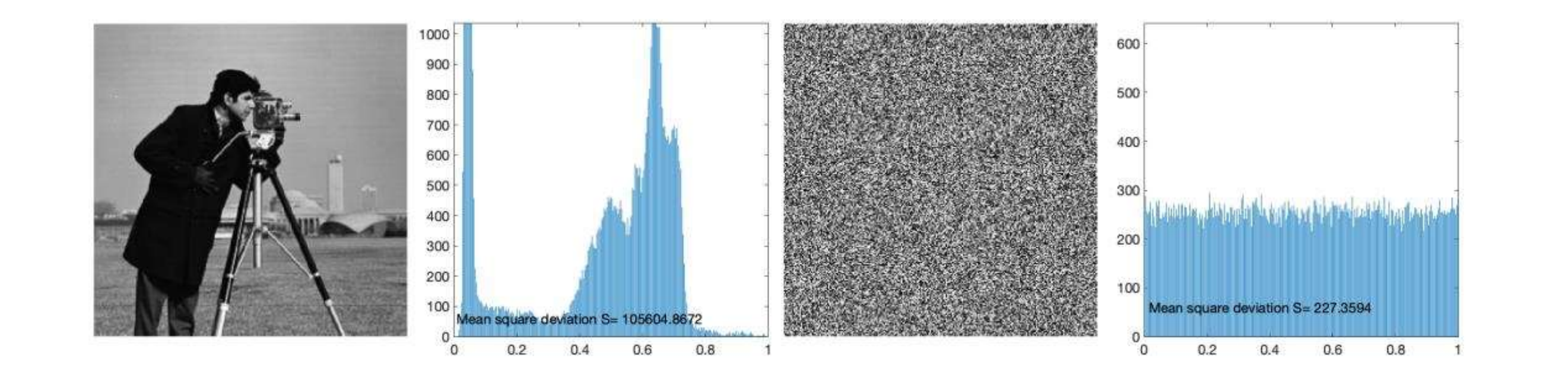}
\includegraphics[width=0.9\textwidth]{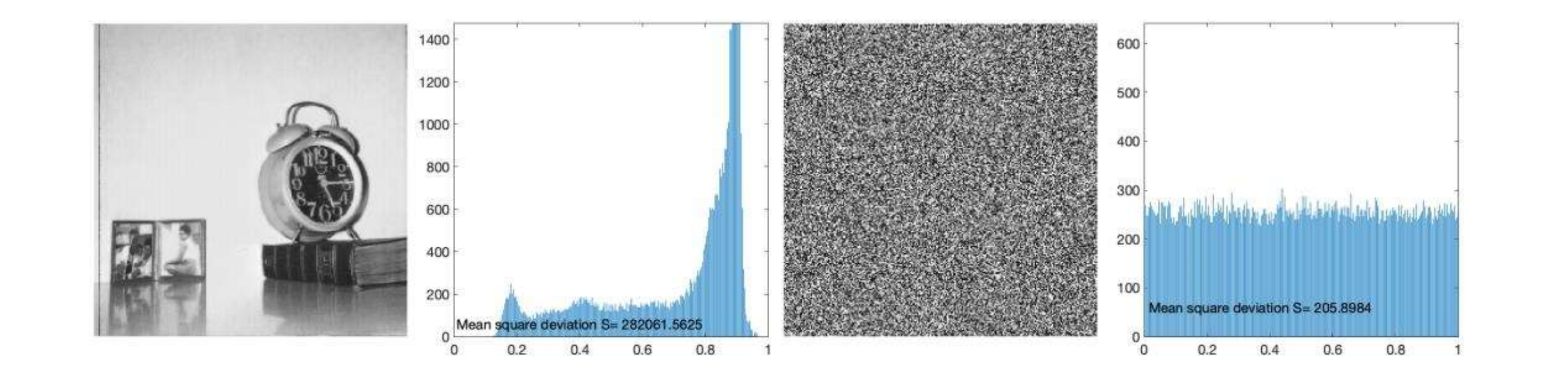}
\includegraphics[width=0.9\textwidth]{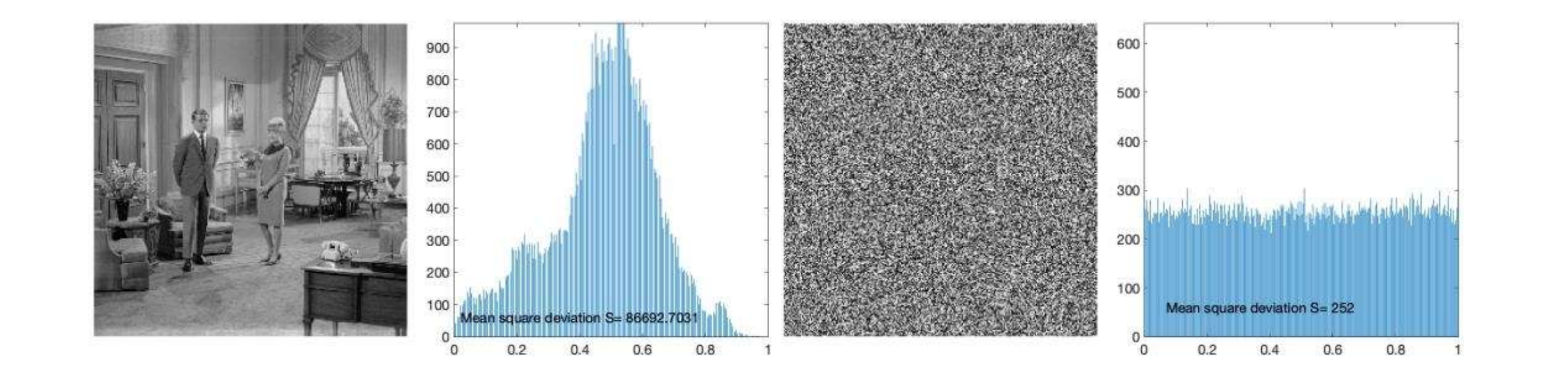}
\includegraphics[width=0.9\textwidth]{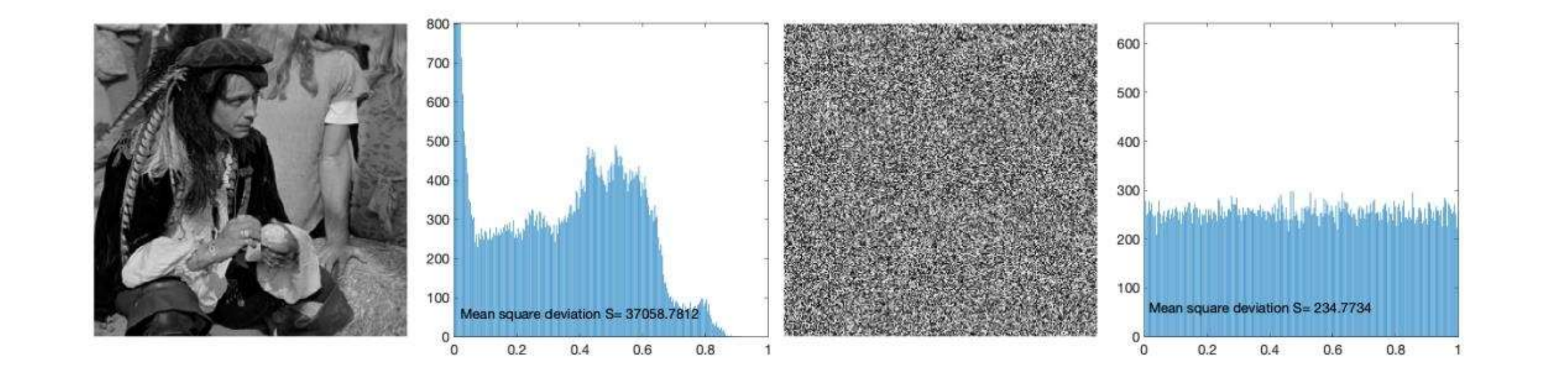}
\setlength{\abovecaptionskip}{0pt}
\caption{Histograms of six images:``Lena, Baboo, Cameraman, Clock, Couple, Man''}
\label{1histogram_whole}
\end{figure*}

Table~\ref{1table_variance} shows the histogram values of six images before and after encryption, and Fig.~\ref{1histogram_whole} shows the histogram distribution of six images before and after encryption. From the data and the figure, we can find that all the histograms of ciphertext images tend to be evenly distributed, and after encryption Lena's variance is as low as 195.766, indicating that this algorithm can effectively resist histogram analysis.

\begin{table}[!htb]
\setlength{\abovecaptionskip}{0pt}
\caption{The variance of six images}
\label{1table_variance}
\centering
\scalebox{0.75}{
\begin{tabular} {p{3cm}p{3cm}p{3cm}}
\toprule
Image & Plaintext & Ciphertext \\
\midrule
 Lena&41398.1016& 195.7656 \\
Baboo& 46866.8281& 238.9062 \\
Cameraman& 105604.8672 &227.3594\\
Clock & 282061.5625& 205.8984\\
Couple& 86692.7031& 252\\
Man&  37058.7812& 234.7734\\
\bottomrule
\end{tabular}}
\end{table}
\subsubsection{Correlation test}
In a plaintext image, there exist strong correlations among adjacent pixels. To resist statistical analysis, correlation in ciphertext images should be as small as possible \cite{behnia2008correlation}. We randomly select 4000 pairs of neighbouring pixels, including three directions~(horizontal, vertical, and diagonal) to measure the correlations. The required calculation formula is listed in (\ref{1formula_corr}):
\begin{equation}
\label{1formula_corr}
r_{uv}=\frac{cov(u,v)}{\sqrt{D(u)}\sqrt{D(v)}}.
\end{equation}
where
\begin{eqnarray}
\left\{
\begin{aligned}
&cov(u,v)=\frac{1}{N}\sum_{i=1}^{N}(u_{i}-E(u))(v_{i}-E(v))\\
&D(u)=\frac{1}{N}\sum_{i=1}^{N}(u_{i}-E(u))^2\\
&E(u)=\frac{1}{N}\sum_{i=1}^{N}u_{i}
\end{aligned}
\right.
\end{eqnarray}
where $u$ and $v$ represent the grayscale values of two neighbouring pixels in the image.

\begin{figure*}
\label{1corr_whole}
\centering  
\subfigure[]{
\includegraphics[width=0.55\textwidth]{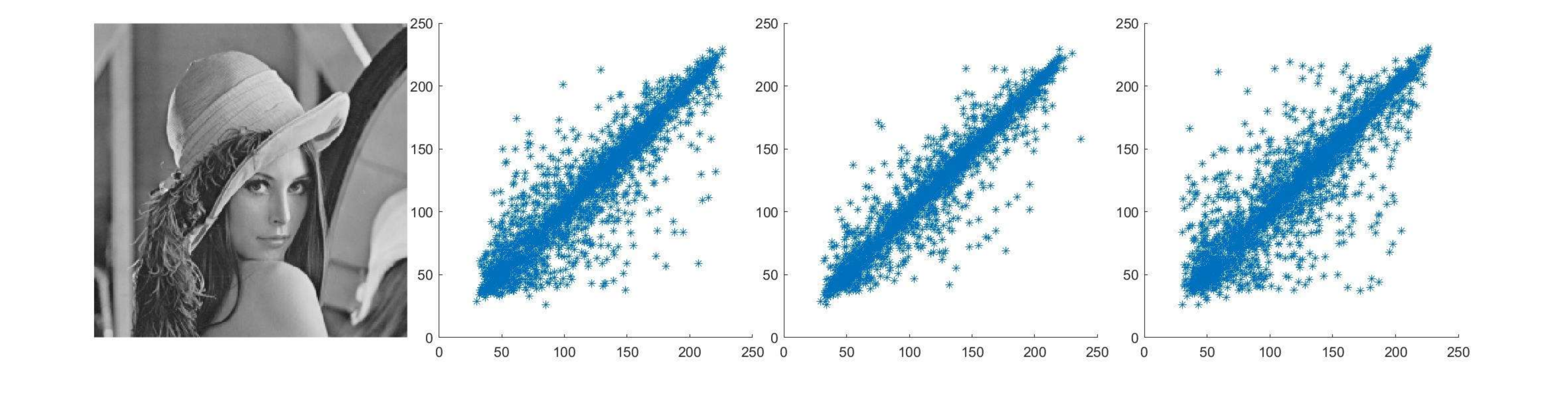}
\hspace{-1.1cm}
\includegraphics[width=0.55\textwidth]{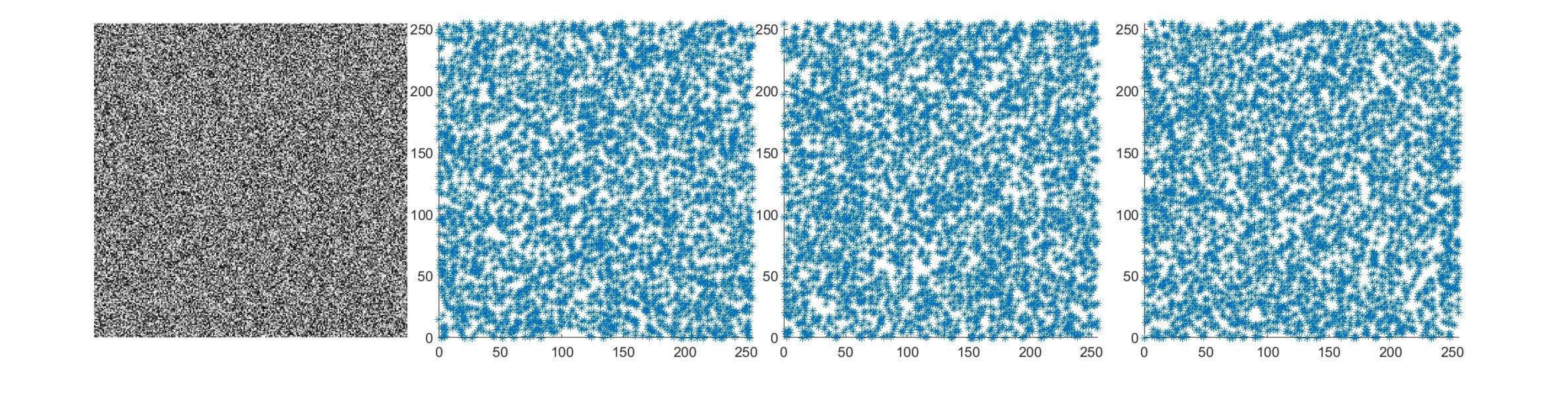}}
\subfigure[]{
\includegraphics[width=0.55\textwidth]{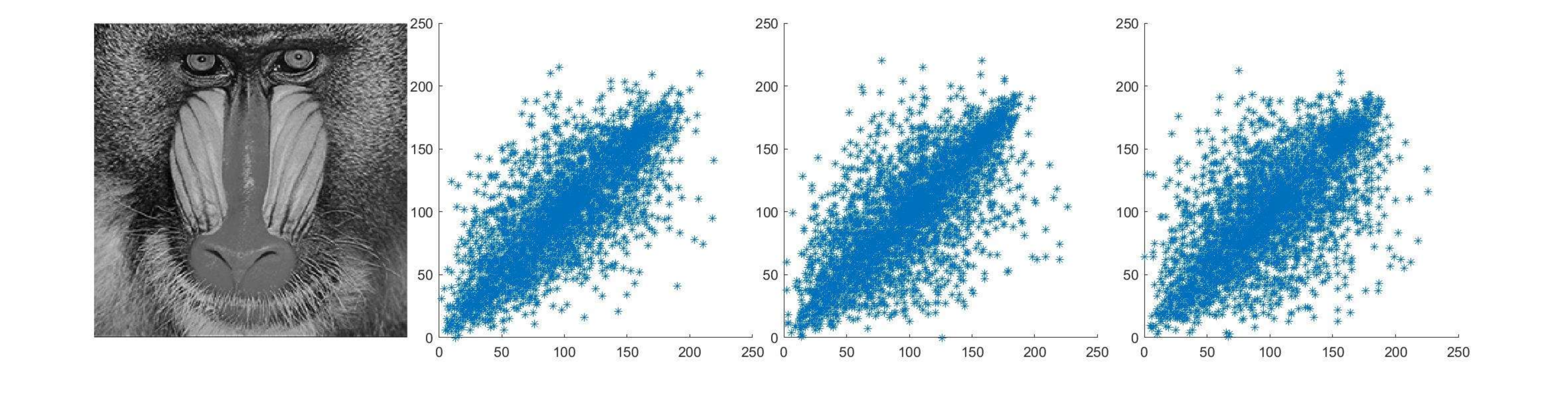}
\hspace{-1.1cm}
\includegraphics[width=0.55\textwidth]{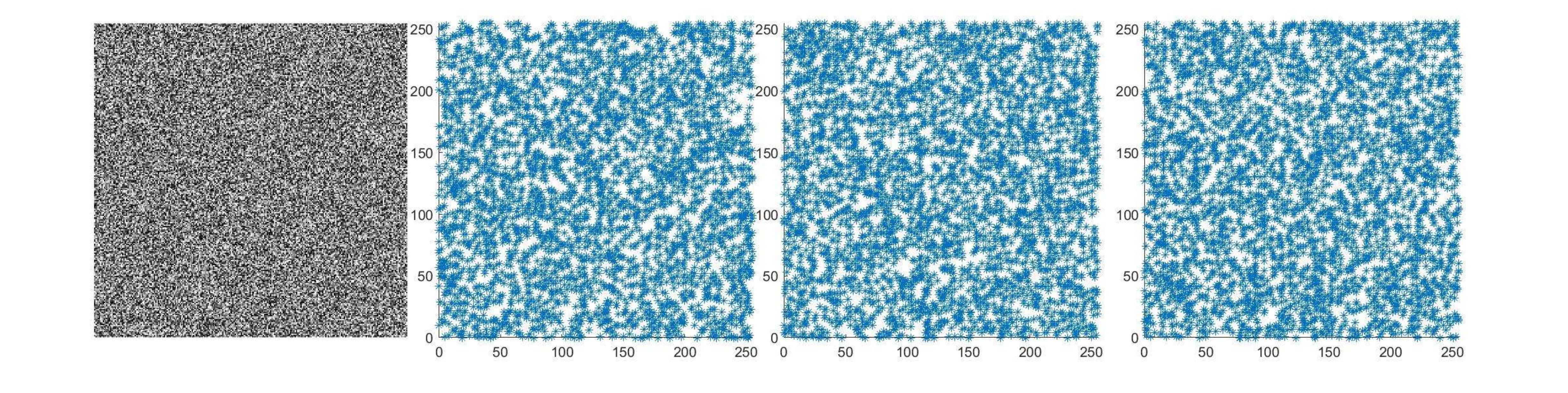}}
\subfigure[]{
\includegraphics[width=0.55\textwidth]{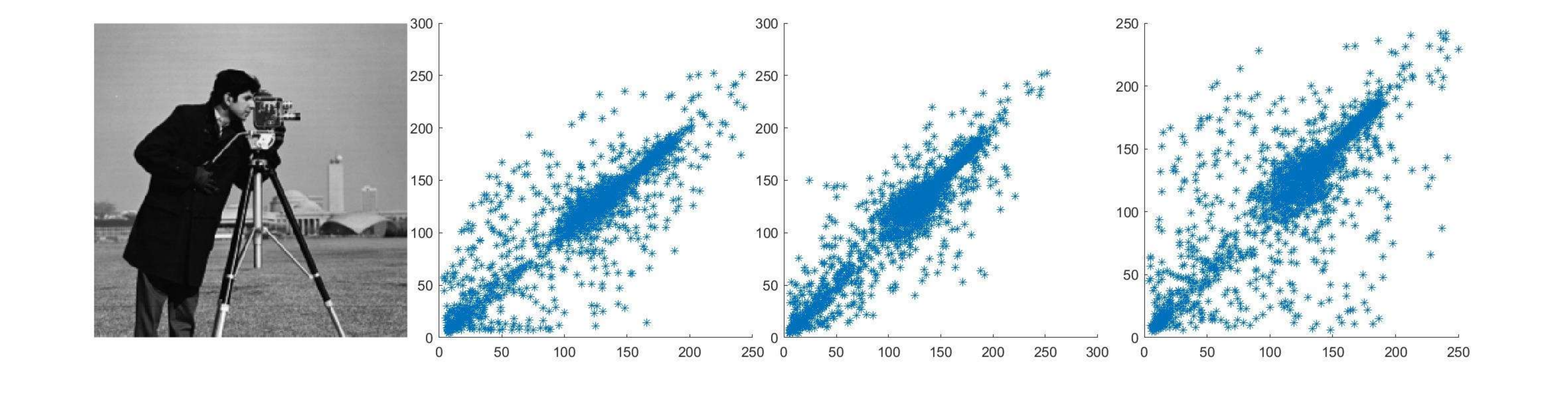}
\hspace{-1.1cm}
\includegraphics[width=0.55\textwidth]{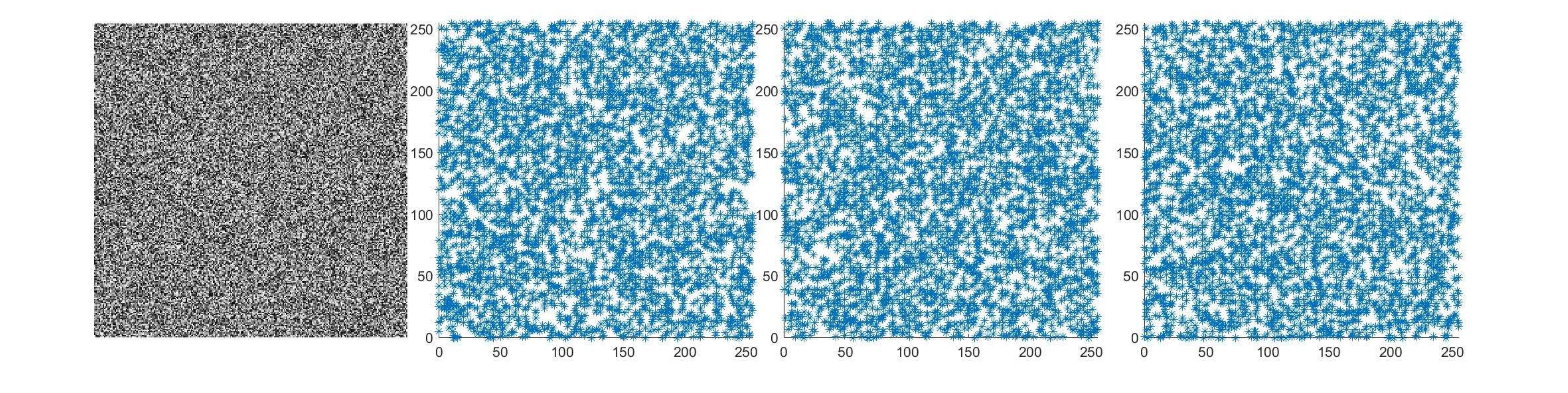}}
\subfigure[]{
\includegraphics[width=0.55\textwidth]{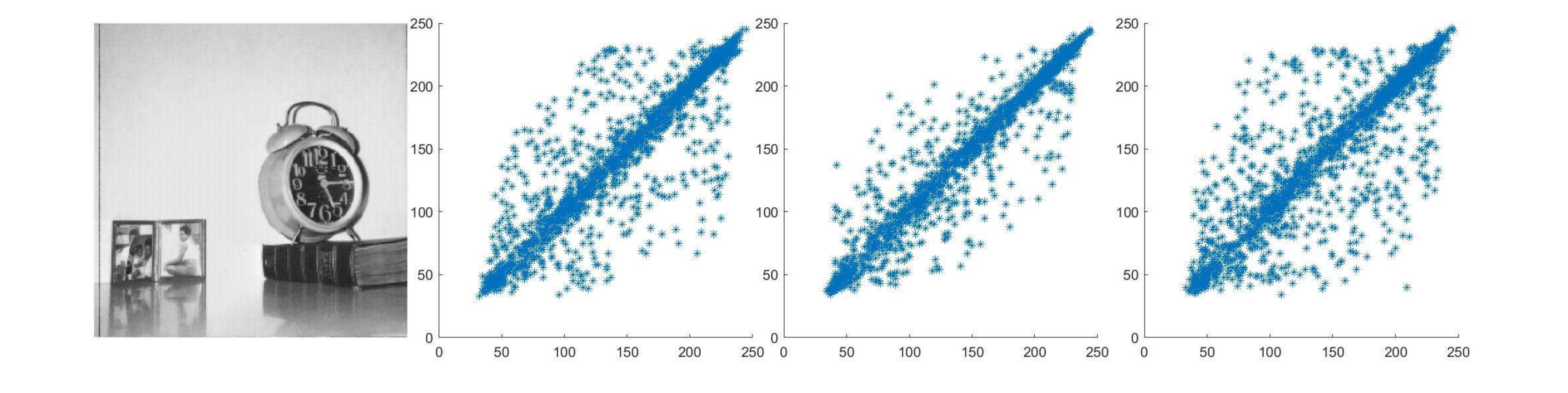}
\hspace{-1.1cm}
\includegraphics[width=0.55\textwidth]{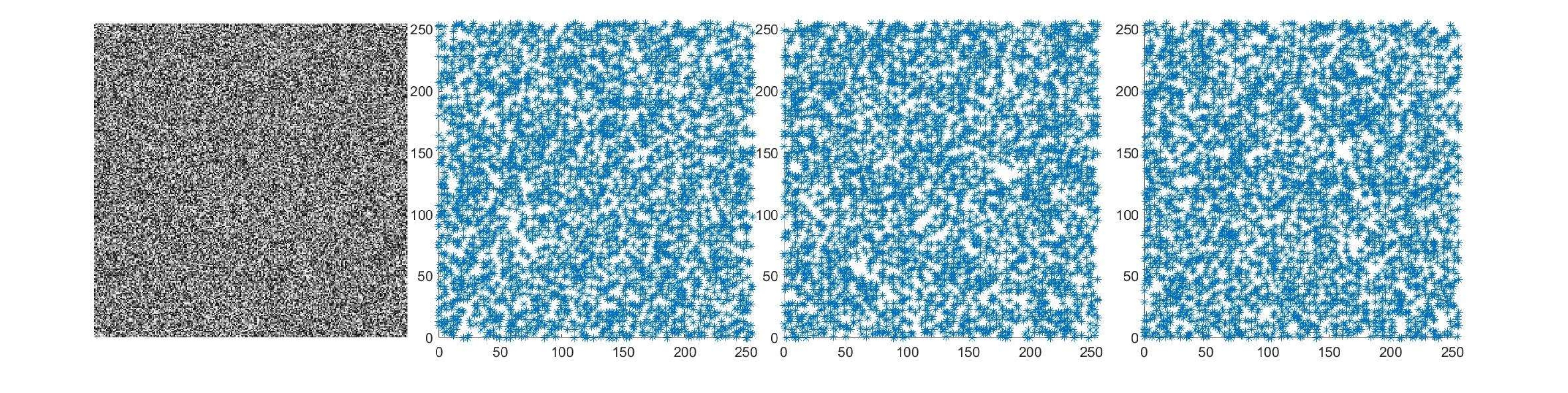}}
\subfigure[]{
\includegraphics[width=0.55\textwidth]{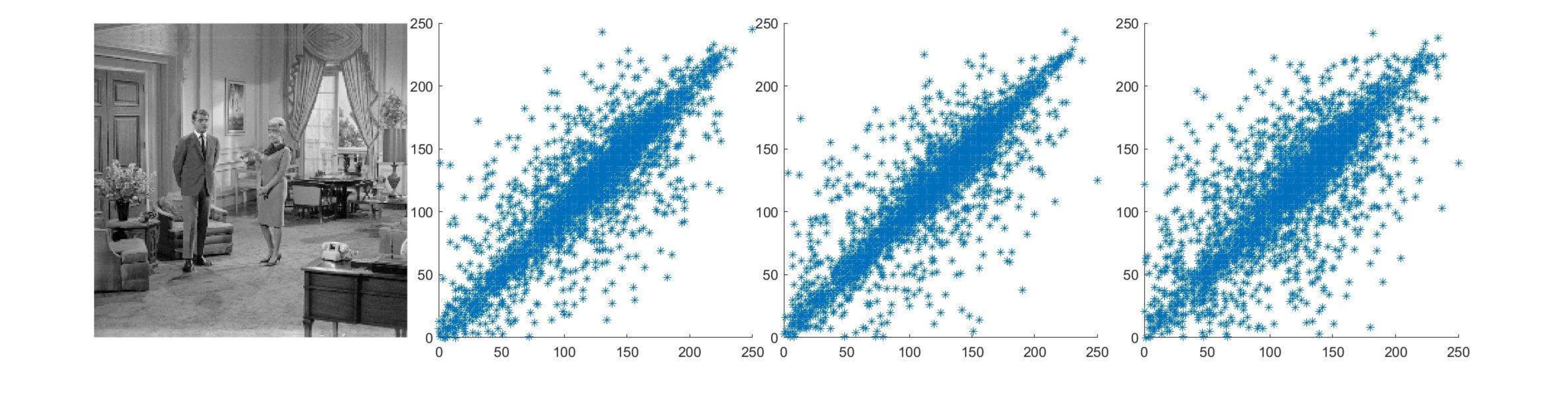}
\hspace{-1.1cm}
\includegraphics[width=0.55\textwidth]{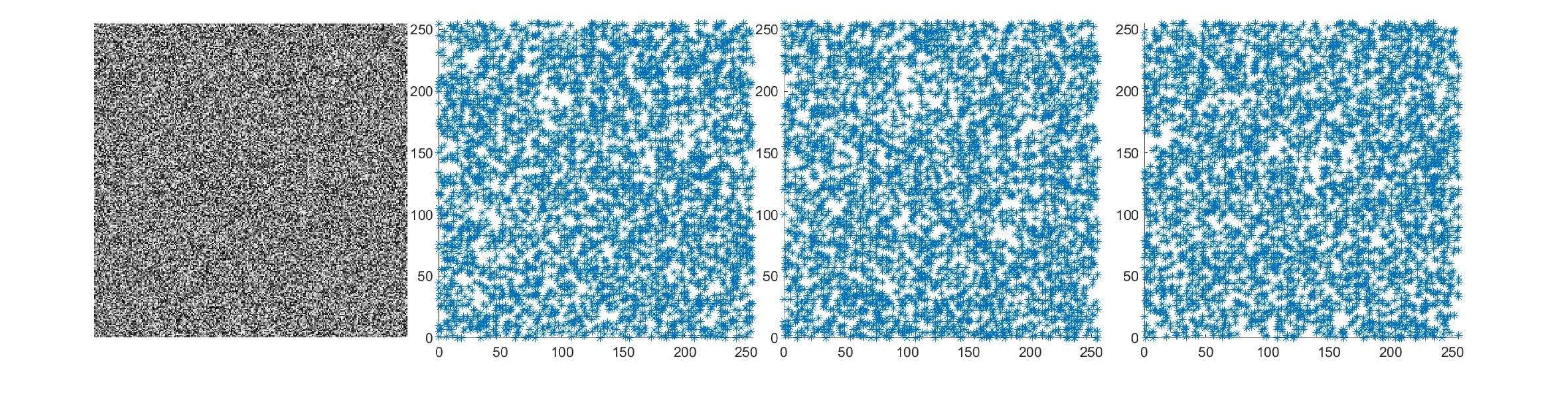}}
\subfigure[]{
\includegraphics[width=0.55\textwidth]{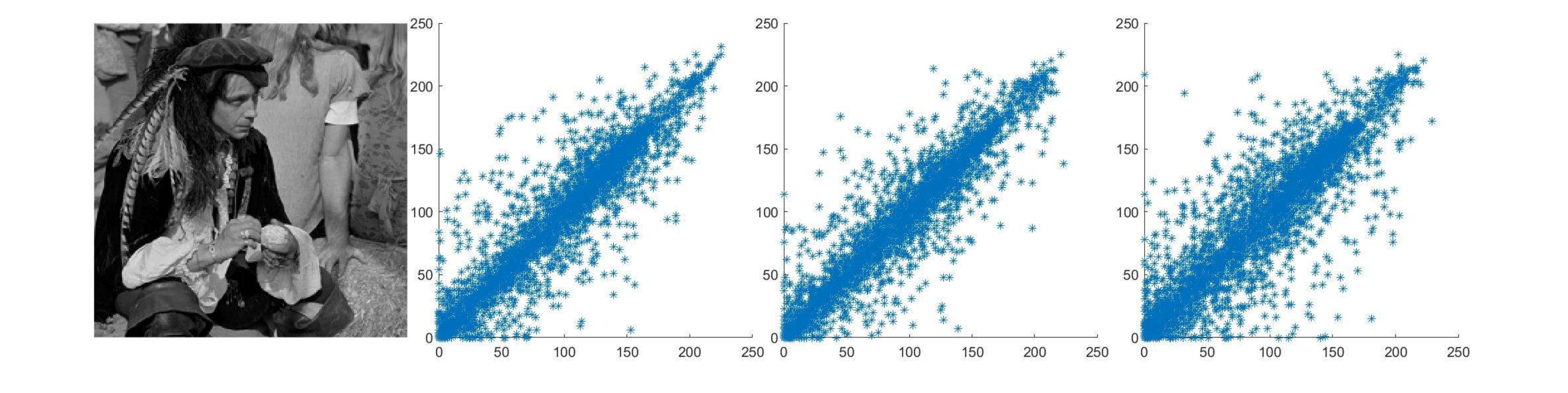}
\hspace{-1.1cm}
\includegraphics[width=0.55\textwidth]{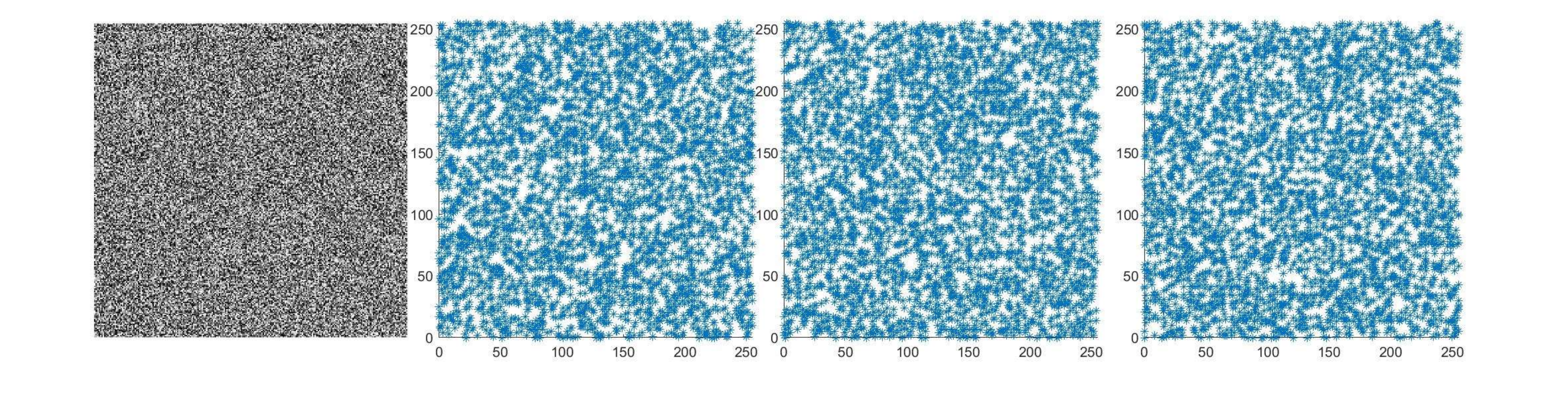}}
\caption{The correlation distribution of plaintext and ciphertext images in the horizontal, vertical and diagonal directions (from left to right) (a)~Lena, (b)~Baboo, (c)~Cameraman, (d)~Clock, (e)~Couple, (f)~Man}
\end{figure*}

To visualize the distribution of images before and after encryption, Fig.~\ref{1corr_whole} displays the correlation distributions of six different images in three directions. Observing the original image we can note that the neighbouring dots are mainly distributed around the diagonal, while in an encrypted image the dots are evenly distributed throughout the whole plane.
That is, the plaintext images are highly correlated in any direction, but the correlations after encryption are very low. Using the calculation formula in Ref.~\cite{xuming2018}, we compute the correlation coefficients of six images before and after encryption, present the results in Table \ref{1table_corr}. We can see that before encryption the values are very large, but after encryption all numerical results are very small, approximate to 0. For comparison with other algorithms, Table \ref{1table_compare_Lena} lists the comparison results in the case of Lena. Although the value of Lena in this article is inferior to Refs.~\cite{liuhui2019Qua4}, it is better than other six Refs \cite{wangxy2021image,xuming2018,xu2017,belazi2016,zhangxuncai2021novel,cao2018novel}.
. The accuracy of the decimal point is $10^{-3}$, which implies that this algorithm achieves a good confusion effect and has strong applicability.

\begin{table}[!htb]
\setlength\tabcolsep{2.5pt}
\centering
\caption{Correlation test results of six images in this article}
\label{1table_corr}
\scalebox{0.7}{
\begin{tabular}{lp{1.3cm}cccc}
\toprule
\multirow{2}{*}{Image} & \multicolumn{3}{c}{Testing direction} & \multirow{2}{*}{Average value} \\
\cmidrule(r){2-4}
&  Horizontal & Vertical &   Diagonal \\
\midrule
Lena    &0.94034 	&0.97136 	&0.92288	&0.94486\\
Ciphertext image of Lena  &-0.00064 &-0.00356&-0.00157 &0.00192 \\
Baboo         &0.78885 	&0.74049 	&0.68020 	&0.73651  \\
Ciphertext image of Baboo  &-0.00291 	&-0.00005 &0.00402 &0.00233 \\
Cameraman    &0.96099 	&0.97463 	&0.92712 	&0.95425\\
Ciphertext image of Cameraman  &0.00198 	&0.00045 	&0.00202 	&0.00148 \\
Clock             &0.95009 	&0.97750 	&0.93230 	&0.95330 \\
Ciphertext image of Clock   &0.00135 	&0.00365 	&-0.00194 	&0.00231\\
Couple          &0.87446 	&0.88660 	&0.80207 	&0.85438 \\
Ciphertext image of Couple &-0.00067 	&-0.00159 	&-0.00023 	&0.00083  \\
Man  &0.93943 	&0.95108 	&0.91287 &	0.93446  \\
Ciphertext image of Man  &0.00347 	&-0.00098 	&-0.00128 	&0.00191\\
\bottomrule
\end{tabular}}
\end{table}

\begin{table*}[!htb]
\centering
\caption{Comparison with other algorithms}
\label{1table_compare_Lena}
\scalebox{0.75}{
\begin{tabular}{llllllllll}
\toprule
\multirow{2}{*}{Image} & \multicolumn{3}{c}{Testing direction} & \multirow{2}{*}{\tabincell{l}{Average \\~~value} } & \multirow{2}{*}{Entropy}& \multirow{2}{*}{\tabincell{l}{NPCR\\~~(\%)}}& \multirow{2}{*}{\tabincell{l}{UACI\\~~(\%)}}& \multirow{2}{*}{\tabincell{l}{Encryption\\~~~time(s)}}& \multirow{2}{*}{\tabincell{l}{Decryption\\~~~time(s)}}\\
\cmidrule(r){2-4}
&  Horizontal &  Vertical    &   Diagonal \\
\midrule
\tabincell{l}{Ciphertext image in\\the proposed algorithm} &-0.0006 &-0.0036&-0.0016 &0.0019 &7.9978&99.617&33.5426&0.3077& 0.2709 \\
Ciphertext image in \cite{belazi2016}   &-0.0048	&-0.0112	&-0.0045&	 0.0068 &7.9973&99.6228&33.7041&0.095&--\\
Ciphertext image in \cite{xuming2018}   &0.0179 &0.022	&7E-06&	 0.0133&7.9970 &99.6107&33.4232 &0.425&--\\
Ciphertext image in \cite{wangxy2021image}  &0.0018	&0.0016	 &-0.0027 & 0.002&7.9974&99.6095&33.4649&0.2-0.23&0.13-0.17\\
Ciphertext image in \cite{liuhui2019Qua4} &0.000882	&0.000108	&0.000019	 &0.000336&7.9974 &99.6102&33.3915&0.1062&--\\
Ciphertext image in \cite{xu2017}    &-0.0226	&0.0041	&0.0368& 0.02117 &7.9963&99.6100&33.5300&0.613&--\\
Ciphertext image in \cite{zhangxuncai2021novel} &0.0023&0.0158&0.0147&	 0.0583 &--&99.6101&33.4583&0.325&--\\
Ciphertext image in \cite{cao2018novel} &-0.0059&-0.0146&0.0211& 0.0139 &7.9973&99.6100&33.4800&0.3243&--\\
\bottomrule
\end{tabular}}
\end{table*}
\subsubsection{Information entropy analysis}
An important measure of testing randomness is information entropy, usually denoted as $H$, which can be measured according to the following formula (\ref{1formula_entropy}):
\begin{equation}
\label{1formula_entropy}
H(m)=-\sum_{i=0}^{l-1}p(m_{i})log_{2}p(m_{i}).
\end{equation}
where $m_{i}$ is the gray value, and there are $l$ kinds of gray values in an image. $p(m_{i})$ represents the probability of $m_{i}$, and  $\sum_{i=0}^{l-1}p(m_{i})=1$. Generally, an image has 256 gray values. Only when the frequency of each gray level is the same, information entropy $H$ reaches the theoretical ideal value 8 \cite{sun2013}.

We use the formula (\ref{1formula_entropy}) to calculate the entropy values of six images before and after encryption, then list the results in Table \ref{1table_entropy}. From the table we can see all values are very close to 8, which shows a good encryption effect. Especially, the information entropy of Lena reaches 7.99784, strong uncertainty of this algorithm has been indicated. Table \ref{1table_compare_Lena} lists Lena's entropy values in different algorithms. Our results are superior to the other six contrast algorithms. Therefore, the ciphertext images have strong uncertainty and our algorithm can resist entropy attacks.
\begin{table}[!htb]
\setlength{\abovecaptionskip}{0pt}
\caption{Information entropy of six ciphertext images}
\label{1table_entropy}
\centering
\scalebox{0.75}{
\begin{tabular} {p{3.2cm}cp{3.2cm}cp{3.2cm}}
\toprule
Image &Original image & Encrypted image \\
\midrule
Lena &7.42489& 7.99784 \\
Baboo &7.37811& 7.99737\\
Cameraman & 7.03056& 7.99748\\
Clock  &6.70567& 7.99775\\
Couple  & 7.05625& 7.99723\\
Man& 7.53608& 7.99741 \\
\bottomrule
\end{tabular}}
\end{table}

\subsection{Differential attack analysis}
A good algorithm can resist differential analysis, requiring different plaintext images (even if with only one different pixel) corresponding to significantly different ciphertext images. In general, there are two commonly used criteria for testing resistance to differential attacks, NPCR and UACI. Let $C_{1}=(C^{1}_{i,j})$ and $C_{2}=(C^{2}_{i,j})$ denote two ciphertext images of size $M \times N$, where their plaintext image has only one different pixel. Define binary sequence to the images $C_{1}$ and $C_{2}$:
\begin{eqnarray}
D_{i,j}=\left\{
\begin{aligned}
&0,~~C^{1}_{i,j}=C^{2}_{i,j}\\
&1,~~C^{1}_{i,j}\neq C^{2}_{i,j}
\end{aligned}.
\right.
\end{eqnarray}

Then define NPCR as formula (\ref{1formula_NPCR}), which means the percentage of different pixels of two ciphertext images.
\begin{equation}
\label{1formula_NPCR}
{\rm NPCR}=\frac{\sum_{i=0}^{M-1}\sum_{j=0}^{N-1}D(i,j)}{M \times N}\times 100\%.
\end{equation}

Furthermore, define UACI as formula (\ref{1formula_UACI}), which means the average of the absolute difference of two ciphertext images.

\begin{equation}
\label{1formula_UACI}
{\rm UACI}=\frac{\sum_{i=0}^{M-1}\sum_{j=0}^{N-1} \left |C^{1}_{i,j}-C^{2}_{i,j} \right |}{255 \times M \times N}\times 100\%.
\end{equation}

Table \ref{1table_NPCR_Lena} shows encrypted Lena's NPCR and UACI at four specific positions, from which we can discover that the values are different at different positions. That is to say, these two indicators have randomness. For unity, let's reduce the first pixel at position (0,0) by 1, calculate the values of NPCR and UACI of six images based on formulas (\ref{1formula_NPCR}) and (\ref{1formula_UACI}), list the numerical results in Table \ref{1table_NPCR}.

\begin{table}[htbp]
\scriptsize
\setlength{\abovecaptionskip}{0pt}
\caption{Lena's NPCR and UACI at specific positions}
\label{1table_NPCR_Lena}
\centering
\scriptsize
\begin{tabular}{ccccc}
\toprule
Location& (209,232) & (33,234) & (162,26)& (72,140) \\
\midrule
NPCR &0.996353& 0.996185 & 0.996292& 0.996246 \\
UACI &0.334139& 0.334155 & 0.33416& 0.334107\\
\bottomrule
\end{tabular}
\end{table}

\begin{table}[htbp]
\setlength\tabcolsep{6pt}
\caption{The NPCR and UACI of six images}
\label{1table_NPCR}
\centering
\scalebox{0.75}{
\begin{tabular}{p{2.8cm}p{2.8cm}p{2.8cm}}
\toprule
Image &NPCR &UACI\\%
\midrule
Lena&0.99617 &0.335426\\
Baboo & 0.996307& 0.334622 \\
Cameraman& 0.996292& 0.332594\\
Clock& 0.996124& 0.3351\\
Couple& 0.996307& 0.335925\\
Man& 0.996078& 0.333822\\
\bottomrule
\end{tabular}}
\end{table}
With a significance coefficient of 0.05, the ideal NPCR is $99.5693\%$, and UACI is $33.2824\%$ for images of size $256\times256$ \cite{wuyue2011npcr}. The results in Table \ref{1table_NPCR} are all higher than the expected values, which proves that the algorithm in this article effectively passes the differential attack capability test. The comparation with other algorithms in case of Lena can refer to Table~\ref{1table_compare_Lena}.

\subsection{Robustness test}
In the process of transmitting ciphertext images over the network, the data may be lost or attacked by noise, which requires the ciphertext image to have good anti-cutting and anti-noise attack performance. In other words, a good algorithm for image encryption should have robustness \cite{wangxy2019color}. In addition, we can use PSNR to evaluate the quality of the decrypted image and the original image. The larger the value is, the more similar the two images, and the formula~(\ref{1formula_PSNR}) is as follows:
\begin{equation}
\label{1formula_PSNR}
\scriptsize
{\rm PSNR}=10\times log_{10} \frac{M \times N \times 255^{2}}{\sum_{i=0}^{M-1}\sum_{j=0}^{N-1}(P(i,j)-C(i,j))^{2}}.
\end{equation}

\begin{figure*}[h]
\centering  
\subfigure[Cut part of ciphertext]{
\label{1cut_1}
\includegraphics[width=0.9\textwidth]{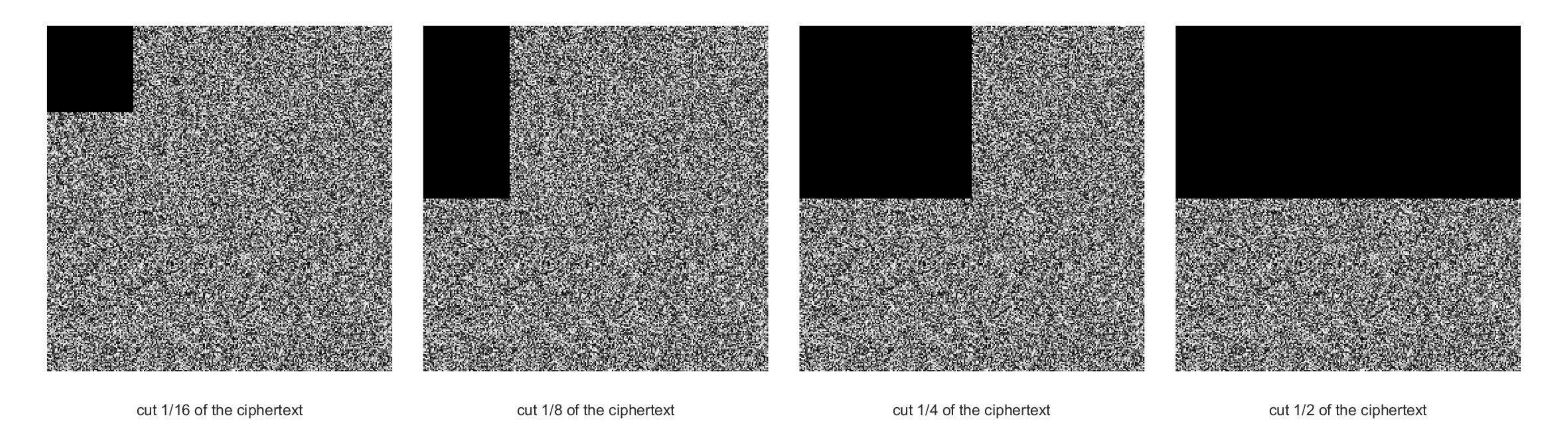}}
\subfigure[Decryption of Fig.\ref{1cut_1}]{
\label{1cut_2}
\includegraphics[width=0.9\textwidth]{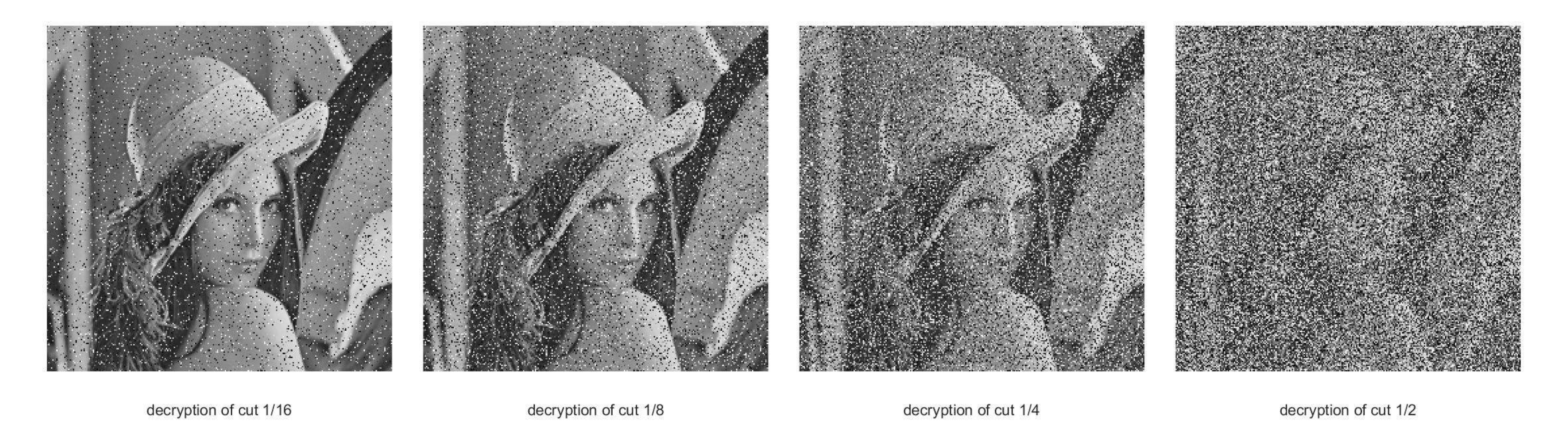}}
\caption{Cutting attack resistance}
\label{1cut_whole}
\end{figure*}

Taking Lena as an example, from the ciphertext image we respectively cut off 1/16, 1/8, 1/4 and 1/2 data at the top left corner, then decrypt the cut ciphertext images using the correct key. Fig.~\ref{1cut_whole} displays the results, which clearly shows that even after its data is cut in half, the body of the image is still visible. The corresponding PSNR values of Fig.~\ref{1cut_whole} are shown in Table~\ref{1table_PSNR_cut}. In summary, this algorithm indicates a good cutting resistance.\

\begin{figure*}[h]
\centering
\subfigure[The encrypted image attacked by different types of noises]{
\label{1noise_1}
\includegraphics[width=0.9\textwidth]{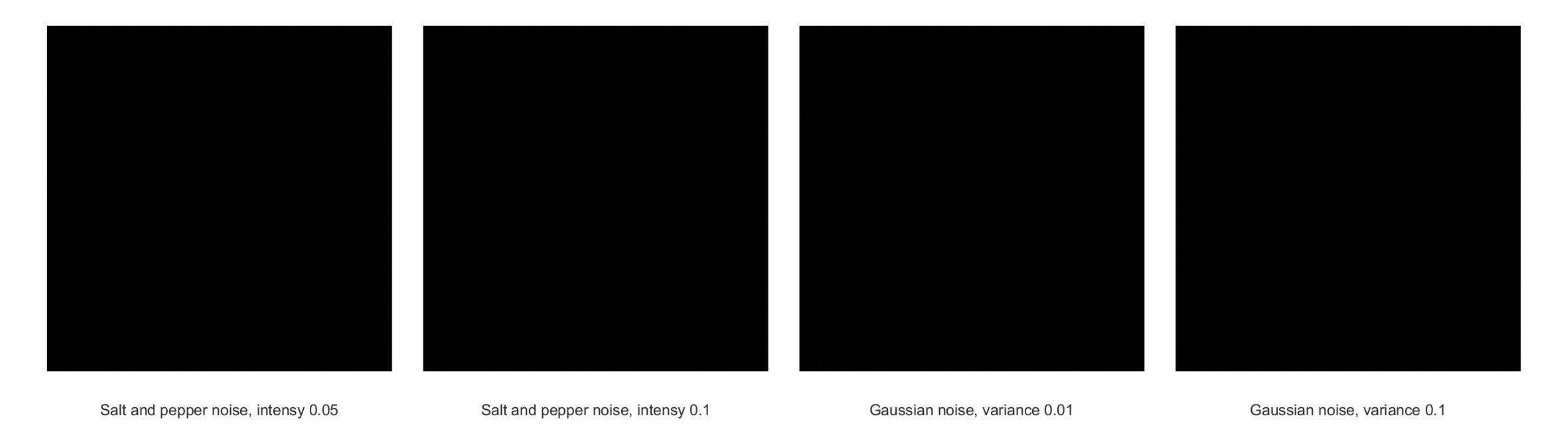}}
\subfigure[Decryption of Fig.\ref{1noise_1}]{
\label{1noise_2}
\includegraphics[width=0.9\textwidth]{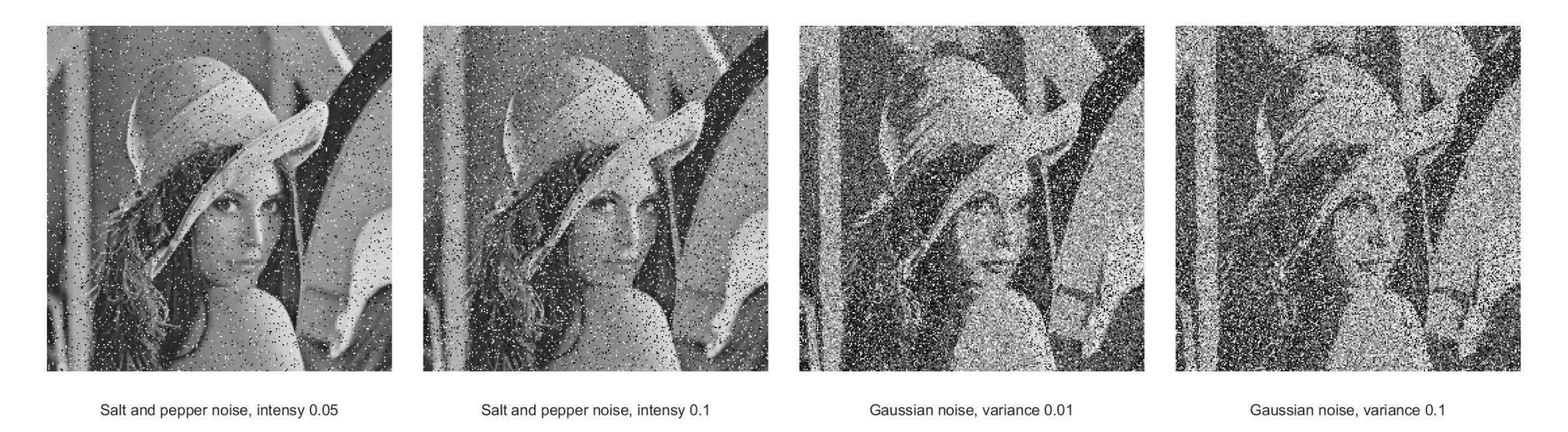}}
\caption{Noise attack resistance}
\label{1noise_whole}
\end{figure*}

Still taking Lena as an example, we respectively use salt and pepper noises with density 0.05, 0.1 and Gaussian noises with variance 0.01, 0.1 to attack. As shown in Fig.~\ref{1noise_whole}, the image is still visible, and the corresponding PSNR values of Fig.~\ref{1noise_whole} are shown in Table~\ref{1table_PSNR_noise}, indicating a good performance to resist noise attacks.

\begin{table}[htbp]
\setlength{\abovecaptionskip}{0pt}
\caption{PSNR with different cutting attacks}
\label{1table_PSNR_cut}
\centering
\scalebox{0.75}{
\begin{tabular}{lcccc}
\toprule
Image & PSNR values(dB)& & & \\
 &\tabincell{l}{cut 1/16}
 &\tabincell{l}{cut 1/8}
 & \tabincell{l}{cut 1/4}
 & \tabincell{l}{cut 1/2}\\
\midrule
Lena  &18.4952 &	15.6102 &	12.7567 &10.3489  \\
Baboo  &18.5505 	&15.5962 &	12.7419 	&10.4122 \\
Cameraman &17.6986 	&14.8126 &	12.0811 &	9.6798 \\
Clock &16.6901 &	13.6156 	&10.7295 &	8.1282 	 \\
 Couple &18.8945 &	15.8033 &	12.9758 &	10.5872 \\
 Man &17.7603 	&14.6126 &	11.7237 	&9.4005  \\
\bottomrule
\end{tabular}}
\end{table}

\begin{table}[htbp]
\setlength\tabcolsep{4pt}  
\setlength{\abovecaptionskip}{0pt}
\caption{PSNR with different noises attack}
\label{1table_PSNR_noise}
\centering
\scalebox{0.75}{
\begin{tabular}{p{1.2cm}cp{2cm}cp{1cm}cc}%
\toprule
Image & PSNR values(dB)& & & \\
 &\tabincell{l}{Salt \& Pepper\\noise(0.05)}
 &\tabincell{l}{Salt \& Pepper\\noise(0.1)}
 & \tabincell{l}{Gaussian \\noise(0.01)}
 & \tabincell{l}{Gaussian \\noise(0.1)}\\
\midrule
Lena  &19.3543 &	16.5183 &	13.2037 &11.9735 \\
Baboo  &19.3658 	&16.2732 	&13.1195 	&11.9209\\
Cameraman &18.4388 	&15.6301 &	12.2959 &	11.1939\\
Clock &17.4574 &	14.4200 &	11.0870 	&9.9733 \\
 Couple &19.4215 &	16.4731 &	13.0921 &	12.0530 \\
 Man &18.1337 	&15.3696 &	11.7741 &	10.7375\\
\bottomrule
\end{tabular}}
\end{table}
\subsection{Computational complexity and time efficiency}
Any effective image encryption algorithm requires low computational complexity. In Algorithm 1, one chaotic sequence is generated with computational complexity O$(n)$, and three Latin squares are constructed with computational complexity O$(3n^{2})$. In Algorithm 2, there is a three-layer encryption structure, which mainly uses an $n$-transversal for the first scrambling and Latin square for auxiliary diffusion and the second scrambling. The computational complexity is O$(n^{3})$, so the computational complexity of this algorithm is O$(n^{3})$.

The fast encryption speed of the algorithm can meet the requirements of instant encryption. The experimental environment is MATLAB R2019b, Microsoft Windows 10 with Intel core i5-1135G7, 2.40 GHz processor and 16 GB RAM.  Table~\ref{1table_time} shows the encryption time and decryption time of six images, which are all calculated 20 times on average. It can be found that all encryption times approximately equal to 0.31s, decryption times approximately equal to 0.27s, and the comparison results of Lena and other algorithms are listed in the Table \ref{1table_compare_Lena}. It can be seen that the algorithm in this paper has relatively fast encryption speed.

\begin{table}[htbp]
\setlength{\abovecaptionskip}{0pt}
\caption{Encryption and decryption time of six images}
\label{1table_time}
\centering
\scalebox{0.75}{
\begin{tabular} {p{2.8cm}cp{2.8cm}cp{2.8cm}}
\toprule
Image &Encryption time & Decryption time \\
\midrule
Lena &0.30769& 0.27087 \\
Baboo &0.30816& 0.27063\\
Cameraman & 0.30946& 0.27259\\
Clock  &0.30812& 0.27233\\
Couple  & 0.30727& 0.27219\\
Man& 0.30769& 0.27266 \\
\bottomrule
\end{tabular}}
\end{table}
\subsection{Resistance to classical types of attacks}
There are four classical types of attacks: ciphertext only, known plaintext, chosen plaintext, chosen ciphertext. Among them, chosen plaintext attack is the most powerful attack. If an algorithm can resist this attack, it can
resist others \cite{wangxy2012novel,zhoujie2020fast}.

The proposed algorithm only needs one round of encryption to achieve a safe effect. It depends on the plaintext image and is very sensitive to the initial parameters $\mu_0$ and initial values $key_0$, $key_1$. If the plain image or one key has changed, $M, M_1$, $M_\gamma$ and $D$ would be totally different. Further more, in the diffusion stage, encrypted value is not only related to plain value and former ciphered value, but also related to the second chaotic sequence and two Latin squares. So, the proposed algorithm can resist the chosen plaintext/ciphertext attack as well as other types of attacks.

\section{Conclusion}
\label{sec5}
In this article, a new chaotic algorithm for image encryption is proposed, mainly using transversals in a Latin square. An $n$-transversal can group all the positions and provide a pair of orthogonal Latin squares. So we can do scrambling in two ways, group by group and using the orthogonality of a pair of Latin squares, which obtains a nice scrambling effect. In the diffusion process, two suitable and uniform Latin squares are selected to do auxiliary diffusion based on a chaotic sequence, achieving good diffusion results. The proposed algorithm has successfully encrypted all test images, and passes the key sensitivity test, statistical test, plaintext sensitivity test, robustness test and other tests. Moreover, the entropy of each encrypted image is very close to 8, the correlation coefficient is very small, close to 0. From the above analysis, the algorithm in this article performs better than other contrast encryption algorithms. It shows that the algorithm achieves a secure and fast encryption effect, simultaneously has robustness and practicability.

\section*{Acknowledgments}
The authors are grateful to Professor Jianguo Lei for some helpful discussions on the subject. This work was supported by National Natural Science Foundation of China~[grant numbers 11871019, 11771119, 61703149]; and Foundation of Hebei Education Department of China~[grant number QN2019127].


\appendix
\section{Proof of Theorem \ref{th1.2.2} }
Proof: 1) let $F=\{g_{0},g_{1},...,g_{n-1}\}$ be a finite field with character $p$. Cayley table $M$ is a Latin square, the ($i,j$)th entry is  $g_{i}+g_{j}$. By the definition of $\gamma_{j}$, it's easy to see that  these $\gamma _{j}$s ($j=0,1,...,n-1$) are $n$ different bijections.\

For any $x\in F$, define the mapping $\sigma _{j}:x\rightarrow x+\gamma _{j}(x)$, $j=0,1,...,n-1$. Then
\begin{equation}
\sigma_{j}(x)=x+\gamma _{j}(x)=x+(ax+g_{j})=(1+a)x+g_{j}.
\end{equation}
Obviously, $\sigma_{j}$ is bijective, then $\gamma _{j}$ is a complete mapping of $F$ under addition. By the definition of $\gamma_{j}$, these $\gamma_{j}$s are $n$ different complete mappings of $F$.

2) By the definition of $M_{\gamma}$, it's easy to see that each element of $F$ occurs exactly once in each row and column of $M_{\gamma}$. So $M_{\gamma}$ is a Latin square too.

3) According to Theorem \ref{th1.2.1} and the definition of $D$, all columns of $D$ form $n$ disjoint transversals of $M$.

Firstly let's prove $M_{1}$ is orthogonal to $M$. By the definition of $M_{1}$, the ($i,j$)th entry is $g_{i}+\gamma_{j}(g_{i})$. That is
\begin{equation}
g_{i}+\gamma_{j}(g_{i})=g_{i}+(ag_{i}+g_{j})=(1+a)g_{i}+g_{j}.
\end{equation}
 Because $a \neq 0,1$ and $a\not\equiv -1$(mod~$p$), $M_{1}$ is still a Latin square different from $M,M_{\gamma}$.

Assuming $M_{1}$ is not orthogonal to $M$, there must exist two positions $(i,j),~(i',j')$, where $(i,j)\neq (i',j')$, such that $(g_{i}+g_{j},(1+a)g_{i}+g_{j})=(g_{i'}+g_{j'},(1+a)g_{i'}+g_{j'})$,
namely
\begin{eqnarray}
\left\{
\begin{aligned}
&g_{i}+g_{j}=g_{i'}+g_{j'}\\
&(1+a)g_{i}+g_{j}=(1+a)g_{i'}+g_{j'}
\end{aligned}
\right.
\end{eqnarray}
Then either $(i,j)=(i',j')$ or $a\equiv -1$(mod~$p$). Whatever either case, it's a contradiction with the definition of $a$ or the assumption. Therefore $M_{1}$ is orthogonal to $M$.

Similarly, we can prove  $M_{1}$ and $M_{\gamma}$, $M$ and $M_{\gamma}$ are pairwise orthogonal Latin
squares. Theorem \ref{th1.2.2} is proved.

\end{document}